## Large area exciton funneling in suspended monolayer $WSe_2$

*Nilanjan Basu [1], Svyatoslav Kostyukovets [1], Mousri Paul [1], Moshe G. Harats *[1,2]*

*[1]Department of Materials Engineering, Ben-Gurion University of the Negev, Be'er Sheva 8410501, Israel*

*[2]Ilse Katz Institute for Nanoscale Science and Technology, Ben-Gurion University of the Negev, Beer-Sheva 8410501, Israel*

** E-mail: mharats@bgu.ac.il*

### Abstract

Exciton funneling is a strain-engineering method which is useful in various applications such as single photon emission, solar cells, and excitonic circuits. One of the major bottlenecks in realizing efficient funneling is the scarcity of configurations that give rise to funneling over a considerable large area. In addition, various works have claimed to observe funneling (on a small length scale) where others claimed to observe a more dominant effect of mechanical gating – exciton-to-trion conversion. To resolve both observations, we used in this work suspended membranes in $10\mu m$ through-holes in SiN membranes and achieved an annular region of funneling with a record area of $\sim 42\mu m^2$ at room temperature - more than an order of magnitude increase than any previous report. We observe exciton-to-trion conversion as well but in a much smaller and confined area. We present how in large area through-holes air pressure plays a crucial role and creates an asymmetric strain gradient which gives rise to the large area funneling. Our work elucidates the role of strain versus strain-gradient in asymmetrical strain profiles which generates both phenomena – exciton funneling and exciton-to-trion conversion – within the same framework.

### Main

Suspended 2-dimensional van-der-Waals monolayers are heavily researched as many physical properties are revealed while these materials are suspended rather than supported. This includes studies such as thermal properties[1], quantum hall effect[2], and charge carriers mobilites[3,4]. Most published works rely on materials suspended over cavities perforated in $SiO_2$/Si substrates which is the standard substrate for nanofabrication of 2-dimensional materials devices. This study compares the usual suspended 2-dimensional materials over pits with a different architecture that is less used - a hole perforated in a SiN membrane. Both geometries are used to study exciton funneling in large area monolayer transition-metal dichalcogenides (TMDs) - $WSe_2$.

Excitons are neutral particles and therefore insensitive to electric fields. A different method is required to transport excitons over large distances. Funneling has been devised as such a method. It is a phenomenon where due to bandgap modulation excitons migrate from a region of large bandgap to a region of low bandgap where mechanical strain is the most convenient method to induce the changes in the bandgap[5]. In $WSe_2$ strain modulation has important applications in emerging fields such as single photon emitters (SPE)[6], excitonic devices[7], and solar cells[8]. If one takes the spatial variation of the excitonic band gap energy $E_g(r)$, then the net force that acts over the excitons is:

$$\boldsymbol{F} = -\boldsymbol{\nabla} E_g \quad (1)$$

One of the easiest ways to spatially vary the $E_g$ is by applying strain ($\varepsilon$) which leads to:

$$\boldsymbol{F} = -\frac{\partial E_g}{\partial \varepsilon}\, \boldsymbol{\nabla}\varepsilon \quad (2)$$

Therefore rather than the strain $\varepsilon$ itself, the strain gradient (SG) $\nabla\varepsilon$ is the relevant physical quantity for efficient funneling[5]. High magnitude SG facilitates the drift of excitons in a particular direction rather than random diffusion of excitons[9,10].

Other methods that are not strain-dependent can modulate the band structure such as composite formation[11], moiré potentials[12,13,] and dielectric engineering[14]. Nevertheless, strain engineering methods have an advantage as pristine material composition can be preserved, the system is reversible as TMDs are purely elastic, and SG can be dynamically tuned[9]. Various strain-engineering methods can generate non-uniform strain such as patterned substrates, wrinkles[15], nanobubbles[16] in TMDs, pressurized membranes[17,18], tip-based indentation methods[9], Piezoelectric actuators[19] and stressor thin films[20]. Unfortunately, to date, each one of these methods is limited either by the funneling efficiency itself[9] or by the full area of the funneling due to the competition between the drift length and diffusion length. For example, nanoindentation setups requires expensive and cumbersome setups which poses a risk of puncturing the monolayer, and in the case of pressurized membranes it requires sensitive pumping equipment and can damage the film if not handled carefully[9]. Strain-engineering with wrinkles[21] or nanobubbles[22] lacks reproducibility, and the funneling area is extremely small.

Apart from these experimental issues, perhaps the biggest limitation of all these methods is the inability to produce large area ($> 10\mu m^2$) funneling. All the strain-engineering methods mentioned above can be efficient in producing only highly localized funneling, as the high SG required for funneling can be generated over a scale less than a micron. The localization of the SG is a feature of the symmetry of the strain created with these strain-engineering methods. In contrast, asymmetric strain, as we will present in this paper, frustrates strain relaxation process[23], which gives rise to a high SG over a large area. For the case of nanoindentation strain-engineering, only the localized area near the tip produces funneling[9,24], whereas in wrinkles funneling is confined to the apices of the wrinkles[15,21] and as a result, the exciton funneling is restricted to a localized spatial area. To date, the largest reported funneling area had a span of $\sim 3\mu m^2$ at room temperature (RT) for a 1-dimensional wrinkle template in monolayer $WSe_2$[18]. Thus, a new geometric architecture that is facile and can produce funneling over a large area, can vastly enrich this field.

In this work we fabricated a suspended monolayer of $WSe_2$ over holes spanning several diameters in a SiN membrane, up to $10\mu m$ in diameter. We observed an inverted U-shaped shape of the membrane, with a circular trench at the edge of the hole. This asymmetric shape gives rise to a natural asymmetric strain profile. Photoluminescence (PL), Raman and Atomic-Force-Microscope (AFM) measurements reveal high radial SG, which is necessary for effective funneling. Strain magnitude of $1 - 2$ % is recorded with a SG of $2 - 3\ \%/\mu m$. Effective funneling is observed over a record-breaking annular area of $\sim 42\mu m^2$ at RT, with a maximum enhancement factor of $\sim 30$ times with respect to the unstrained suspended flake at the center. We compared our results to a $10\mu m$ standard pit in $SiO_2$/Si where we did not observe high effective SG - resulting in negligible funneling efficiency as reported in previous works[10,25]. We support our experimental findings with a numerical model that takes into account the role of the asymmetric SG.

## Results

We achieved large area ($\sim 1mm$) monolayer $WSe_2$ flakes by gold-assisted exfoliation over PDMS and dry-transferred the flakes over both holes in a SiN membrane and pits in $SiO_2$/Si[26],[27] (more details in the SI and Methods section). Fig. 1(a-c) and (d-f) shows the schematic, AFM image and height profile of monolayer $WSe_2$ flake over a $10\mu m$ pit and hole, respectively. The measured height profile of the flake over the pit has shown a standard U-shaped profile, dipping approximately $400nm$ at the center of the pit. Observing the height profile of the hole, we see a striking difference of an inverted U-shaped profile. The difference between the height profiles can be explained from the transfer process of the flakes from the PDMS stamp. In the case of the U-shaped profile of the pit, a fraction of the air is squeezed out during the stamping, and as the flake is impermeable, it creates a U-shape even when the PDMS stamp is released due to the balance between the air-pressure from the bottom and top of the flake. The situation is different for the hole, as it is a through-hole and the air pressure balances out both sides while the flake tries to match the initial circumference of the hole when we release the PDMS stamp. Therefore, a circular trench is formed at edge of the hole due to the SiN-flake adhesion at the inner edge of the hole. For comparison, optical micrographs (OM), AFM and SEM images of both pits and holes of 3,5, and $10\mu m$ are shown in the SI. Interestingly, while pits of smaller dimension ($3,5\mu m$) keep the U-shaped profile as well, the smaller diameter holes ($3{,}5\mu m$) did not show an inverted U-shaped profile as observed in the $10\mu m$ case but rather a regular U-shaped profile. The explanation for the difference in the height profile between small ($3{,}5\mu m$) and large ($10\mu m$) holes stem from the balance between the force from the air pressure which is proportional to $r^2$ ($r$ being the hole radius) while the adhesion force between the SiN and flake at the circumference is proportional to $r$. As the hole radius increases, the tendency for the flake to rise to the level of the SiN substrate when the PDMS stamp is released, increases. This leads to the irregular shape of the circular trench that has two different slopes: one coming from the adhesion to the SiN wall, and the other from the rising of the flake. These two slopes give rise to different curvatures, leading to a non-uniform and relatively high magnitude SG which, as we will show experimentally, facilitates the funneling process. On the other hand, the pit (and small holes) has a uniform curvature which naturally gives rise to a uniform strain with low SG, thus no prominent funneling was observed. We note that the gravitational force acting over suspended flakes is negligible and was therefore omitted from our analysis. In addition, we measured the adhesion of the flakes to both substrates (SiN and $SiO_2$) and found it to be similar, excluding adhesion effects related to the suspension shape of the flakes (see SI).

The strain and SG maps of both the 10 $\mu m$ pit and the hole is computed from the AFM data. Geometric strain can be directly calculated from the surface topography by adopting the Green-Lagrange formula (for small strain values) for a line profile in x direction[28]:

$$\varepsilon_x = \sqrt{1 + \left(\frac{dz}{dx}\right)^2} - 1 \tag{3}$$

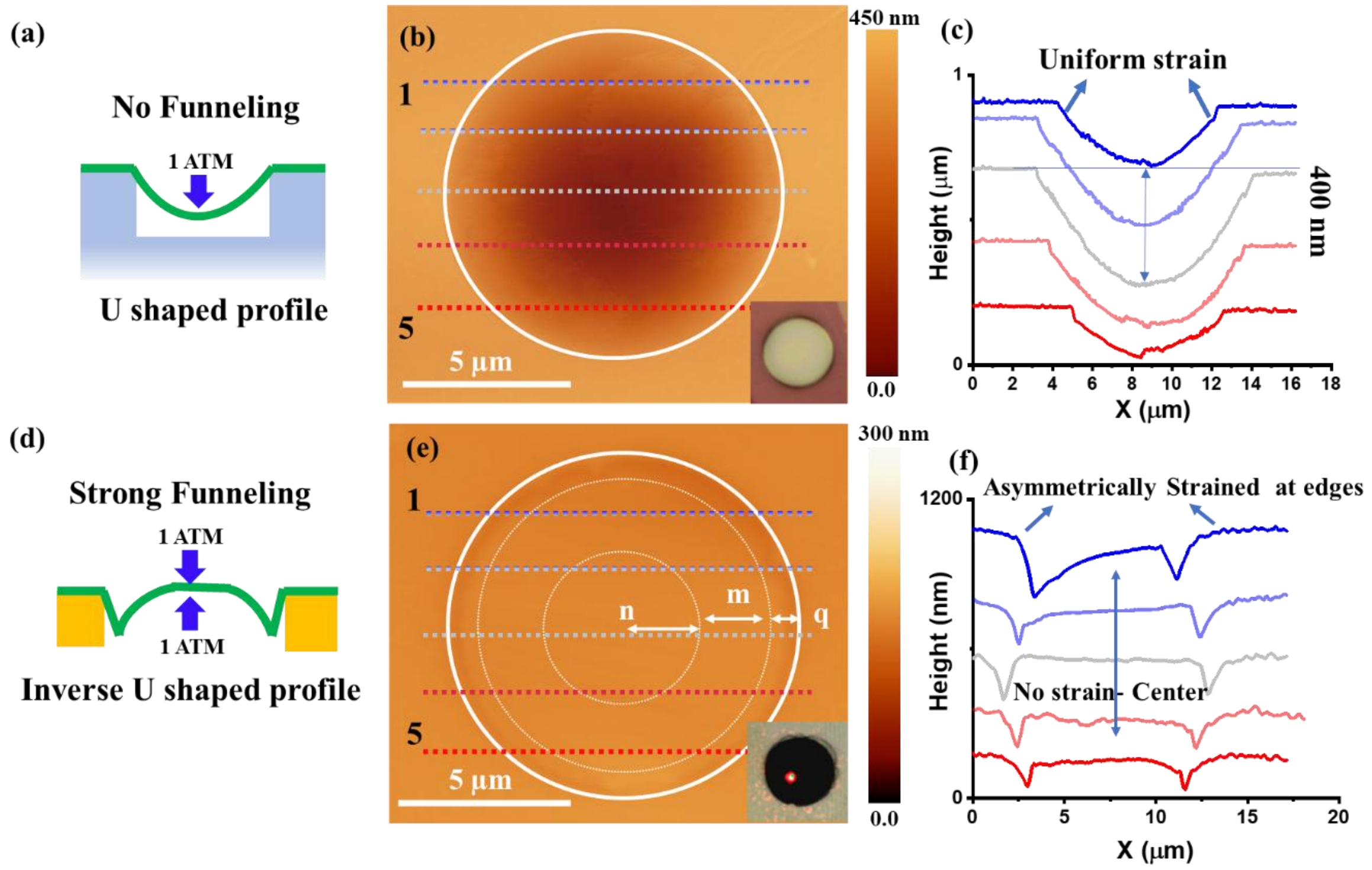

Fig. 1 **Characterization of microscale pits and holes.** (a) Schematic of the cross section of a flake over a pit. The transfer process creates a U-shape profile. (b, c) AFM image and corresponding height profile (shifted for clarity) of a 10 $\mu m$ pit. Inset shows the OM. Thick solid white circle is the circumference of the pit. (d) Schematic of the cross section of a hole where for large holes the transfer process creates an inversed U-shaped profile. (e, f) AFM image and corresponding height profile of the hole shifted for clarity. Inset shows the OM with the reflection of the laser spot. $n, m$, and $q$ are three different regions based on the magnitude of strain and are noted in the main text. Thick solid white circle is the circumference of the hole.

The strain generated by the hole architecture is uniaxial in nature (further details are provided in the SI). Naturally, the pit presents uniaxial strain as well. The strain map of the pit is shown in Fig. 2a which shows ~ 0.7% strain just near the circumference and negligible strain at the center of the pit. The corresponding radial SG map shows a near homogeneous SG with a maximum value of ~ $0.6\%/\mu m$ which is insufficient for funneling. As the strain is uniaxial and varies along the radius, low amplitude concentric patterns can be seen in the radial SG map.

The strain map of the hole is presented in Fig. 2d, which shows maximum strain of ~ 1.3%. Note that the high strain is seen outside the circumference of the hole – presenting the strain from the wall of the SiN – and inside – presenting the inversed U-shape profile. We extract the radial (cylindrical) SG of the hole with respect to its' center in Fig. 2e. The SG map shows a radial distribution of SG as expected from the symmetry of the suspended flake. The high SG area compared to the pit stretches further toward the center of the suspended part rather to the supported region of the flake, which means that the spatial profile is asymmetric as shown in the schematic of Fig. 2f. This differs from previous works where the funneling was symmetric[29,30,31] with respect to the point of high strain. Additional strain maps of different diameters of holes and pits are presented in the SI.

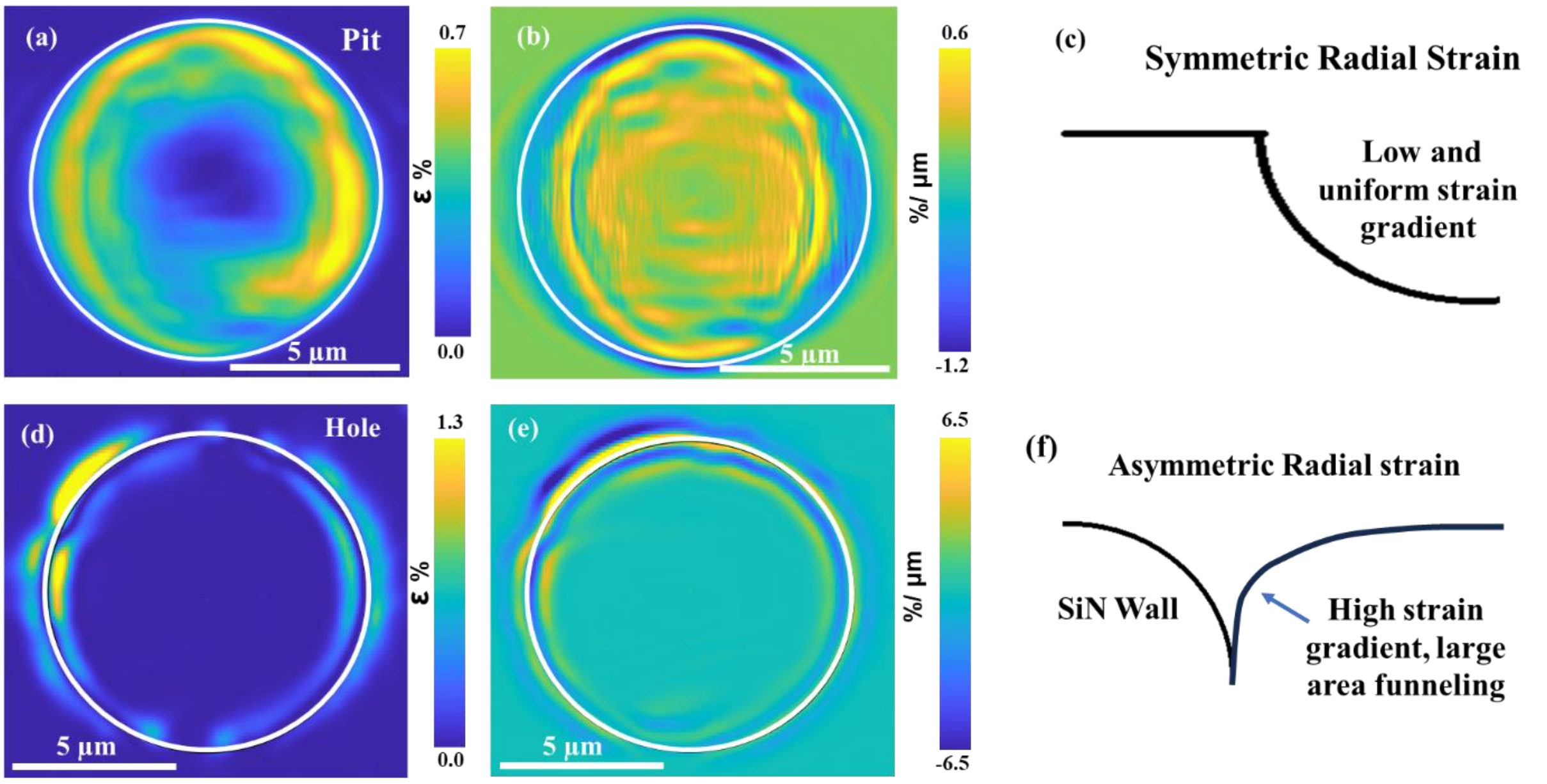

Fig. 2 **Strain and strain gradient from AFM scans.** Solid white circles represent the circumference of the pit/hole (a) Strain map of the $10\mu m$ pit in x-y coordinates. (b, c) Radial strain gradient along $\hat{\mathbf{r}}$ of the $10\mu m$ pit and schematic of the symmetric radial strain respectively. (d) Strain map of the $10\mu m$ hole in x-y coordinates. (e, f) Radial strain gradient along $\hat{\mathbf{r}}$ and schematic of the asymmetric radial strain respectively.

PL measurements of both pits and holes were done at room temperature to investigate the exciton funneling of both geometries. Fig. 3a shows the spatial map of the maximum intensity (raw data) of the PL for the pit, and the corresponding spectra from different locations on the sample (Fig. 3b). Details of the PL measurements are in the Methods section. The intensity over the $SiO_2$ was poor and showed a small red shift (due to screening from the $SiO_2$) w.r.t the center. At the center we observed a slight redshift of about ~ $5\ meV$, corresponding to ~ 0.1% strain w.r.t the circumference. The almost uniform intensity across the pit is an indication that funneling is negligible. The pit exhibits PL brightening, with the brightest part being at the center. The increased brightness compared to the supported part of the flake is a direct result of the difference in refractive indices between the supported and the suspended flake as $SiO_2$, being a high refractive index material (compared to air) guides the light into the substrate[32,33,34].

A similar map for the hole is presented in Fig. 3c with corresponding spectra (Fig. 3d). The center is unstrained (as presented in Fig. 2d) and has low PL intensity. The spectra taken at some distance from the circumference ($\sim 0.8\mu m$) showed a pronounced redshift of $\sim 26\ meV$ and an enhancement factor of $\sim$ 10 times w.r.t to the unstrained part. On one hand, the redshift results from the strain, while on the other hand, the enhancement can be attributed to the drift of excitons due to pure funneling. Further PL maps at different excitation powers and hole size are presented in the SI. Based on the strain profile of the hole that is shown in Fig. 2d, we divide it into 3 distinct regions as depicted in Fig. 3b: region $\boldsymbol{n}$ (radius $\sim 3.2\mu m$ roughly) – center of the hole with negligible strain and no PL enhancement, region $\boldsymbol{m}$ (annular width $\sim 1\mu m$) – between the center and the circumference (near edge) with moderate strain, and region $\boldsymbol{q}$ (annular width $\sim 0.8\mu m$) – the spatial region where the strain is maximum. Fig. 3e,f shows the excitonic energy map and exciton-to-trion ratio for the hole, respectively. The exciton-to-trion conversion is crucial as we want to exclude it from the funneling area is it competes with the funneling as reported previously[9,35]. As the maps of PL intensity are not conclusive evidence for funneling, to estimate the actual funneling area $\boldsymbol{m}$, we incorporated three mutually inclusive conditions – ($i$) enhancement factor ≥ 5, and ($ii$) minimum excitonic shift of 8 $meV$, and excluding the area where ($iii$) exciton-to-trion ratio is higher w.r.t the ratio in the middle of the hole (region $\boldsymbol{q}$). Using these criteria, we find a record funneling area of $41.7^{+2.3}_{-4.0}\ \mu m^2$ as shown in Fig. 3g (the errors result from the errors in defining the exciton-to-trion ratio at the center). We corroborate these findings with additional analysis of the PL spectra and Raman spectroscopy that complete the correlation between the strain gradient and the observed funneling.

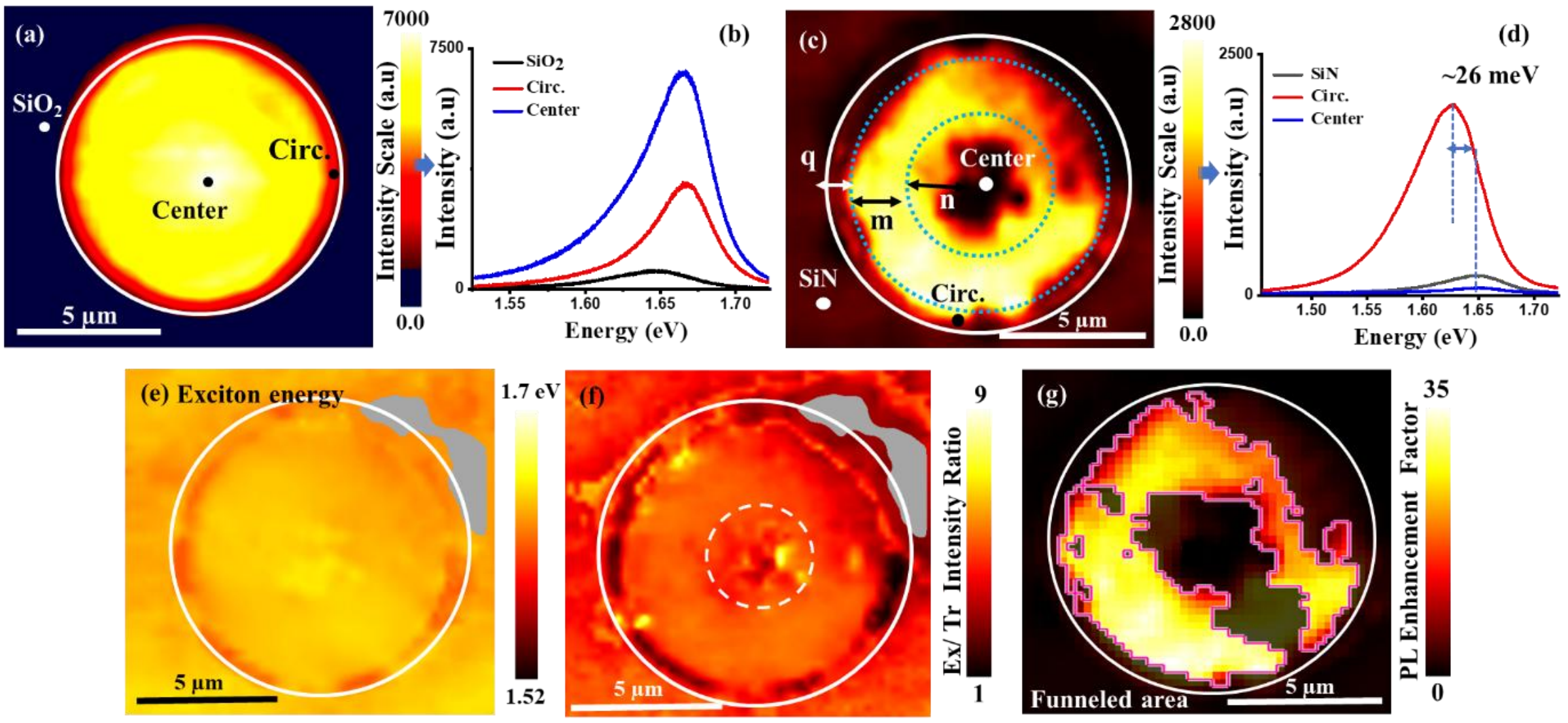

Fig. 3 **PL enhancement through funneling.** (a) Map of the maximum PL intensity for the pit and (b) corresponding spectra from the center (blue), circumference (red), and supported regions of the flake (black). (c) Same map for the hole and (d) corresponding spectra from the center (blue), circumference, (red), and supported regions of the flake (black). (e)-(f) Exciton energy and exciton-to-trion ratio maps of the hole, respectively. Note the efficient exciton-to-trion conversion only at the circumference of the hole (f) at the point of the maximum strain seen from the shift of the exciton energy in (e). (g) Funneling area calculated following the criteria in the main text showing a total funneling area of $41.7^{+2.3}_{-4}\ \mu m^2$.

We derive the strain and SG spatial map from the PL spectra of the pit and hole in Fig. 4. Details of the analysis are presented in the SI. Fig. 4a shows a map of the exciton energy of the pit, showing near uniform exciton energy with a slight variation near the circumference. The shift of the exciton energy was used to generate the strain and radial SG map of the pit as shown in Fig. 4b,c respectively. The maximum strain in the pit is observed at a small distance $\sim 0.5\mu m$ from the circumference, confirming our previous observation from the AFM strain map in Fig. 2a. The SG map shows a rather uniform gradient ($< 0.25\ \%/\mu m$), which explains low funneling efficiency.

The equivalent exciton energy map of the hole is presented in Fig. 3e, which shows a distinct circular red shift pattern along the trench. As with the pit, the shift of the exciton energy was used to generate the strain map from PL data and the SG as well as shown in Fig. 4**Error! Reference source not found.**d,e respectively. As the strain is, for a very good approximation uniaxial in the radial direction, the radial SG map along $\hat{r}$ shows concentric rings. The radial SG $(\frac{\partial \varepsilon}{\partial r})$ increases when we move away from the center and reaches $0.62 \pm 0.02\ \%/\mu m$ just before the trench. At the trench the SG reverses direction $(-\hat{\boldsymbol{r}})$ due to the direction definition of the SG in cylindrical coordinates.

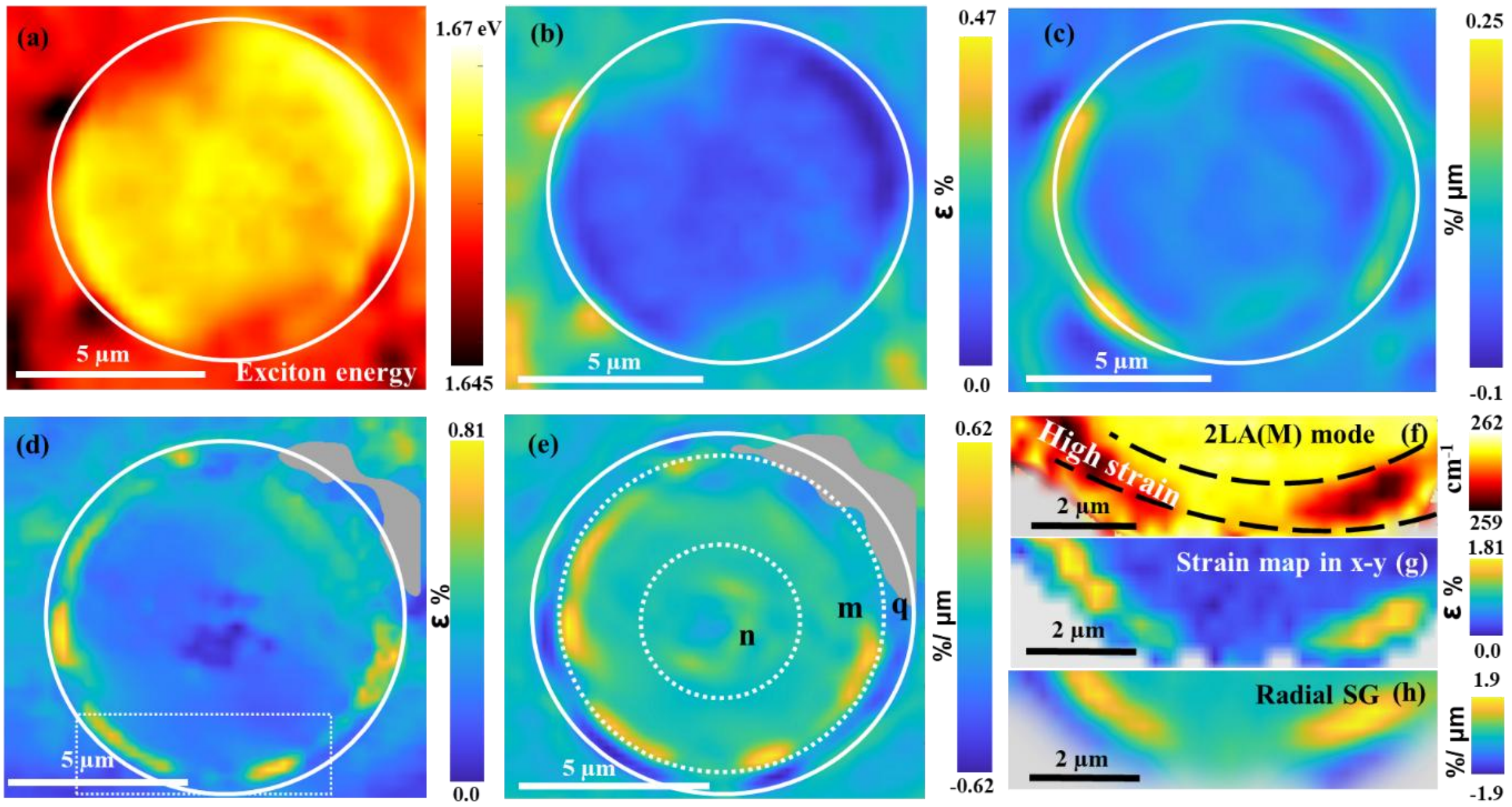

Fig. 4 **Strain and strain gradient from PL and Raman spectra.** (a) Exciton energy map of the pit. (b) Strain map from the exciton map in x-y coordinate and (c) radial SG along $\hat{\boldsymbol{r}}$ of the pit. (d) Strain map from the exciton map (shown in Fig. 3e) in x-y coordinate. The white rectangle indicates the area of the Raman map in (f). (e) radial SG along $\hat{\boldsymbol{r}}$ of the hole. (f) Raman map of the 2LA(M) peak position. (g) Strain map from the 2LA(M) peak position in x-y coordinate and (h) SG in radial $\hat{\boldsymbol{r}}$ direction computed from strain map. Note that although the amplitude of the strain and SG from the Raman is higher than the values derived from the PL, the overall spatial dependency is the same as the area in the rectangle om (d).

Raman measurements were performed on the lower part of the hole to strengthen the picture of the strain and SG in a more accurate manner than PL (see SI for additional Raman data). The 2LA(M) has shown a high shift towards lower wavenumber in the trench region which indicates high strain in the trench[36] as shown in Fig. 4f. Strain and SG maps are generated from the 2LA(M) and shown in Fig. 4g,h, respectively. The maximum strain extracted from the 2LA(M) mode is $1.81 \pm 0.03\ \%$ and the maximum SG is $1.9 \pm 0.2\ \%/\mu m$ (note the lower errors compared to the data obtained from the PL). As in the case of the SG from the PL, the maximal SG is found just beside the trench. We see that the PL strain value is lower by almost a factor of 2 compared to the Raman strain value which shows that ($i$) Raman is a more sensitive method for the determination of the strain value due to the localization of phonons under strain and ($ii$) as PL averages over the inhomogeneous strain due to the processes of funneling and diffusion while Raman is not affected by these physical processes [9]. We conclude that the strain and SG estimations from the Raman data are closer to the values obtained from AFM data and are therefore more accurate for understanding the funneling observed in our samples.

## Discussion

We have shown that with the large, $10\mu m$ hole perforated in SiN, the exciton funneling exceeds an area of $\sim 42\ \mu m^2$ at RT – a feature that has yet to be observed. The experimental data we presented shows that the inquired ingredients for achieving high funneling area stem from large SG over a large area, combined with low mechanical gating. We are able to achieve this with large holes due to the inversed U-shape profile that results in asymmetric strain profile which translates to high SG over an annular ring that is naturally large enough due to the large diameter of the hole. The exciton-to-trion conversion is observed as well but it does not limit the funneling (as seen in previous works)[35] as it is limited spatially only at the point of high strain at the circumference of the hole.

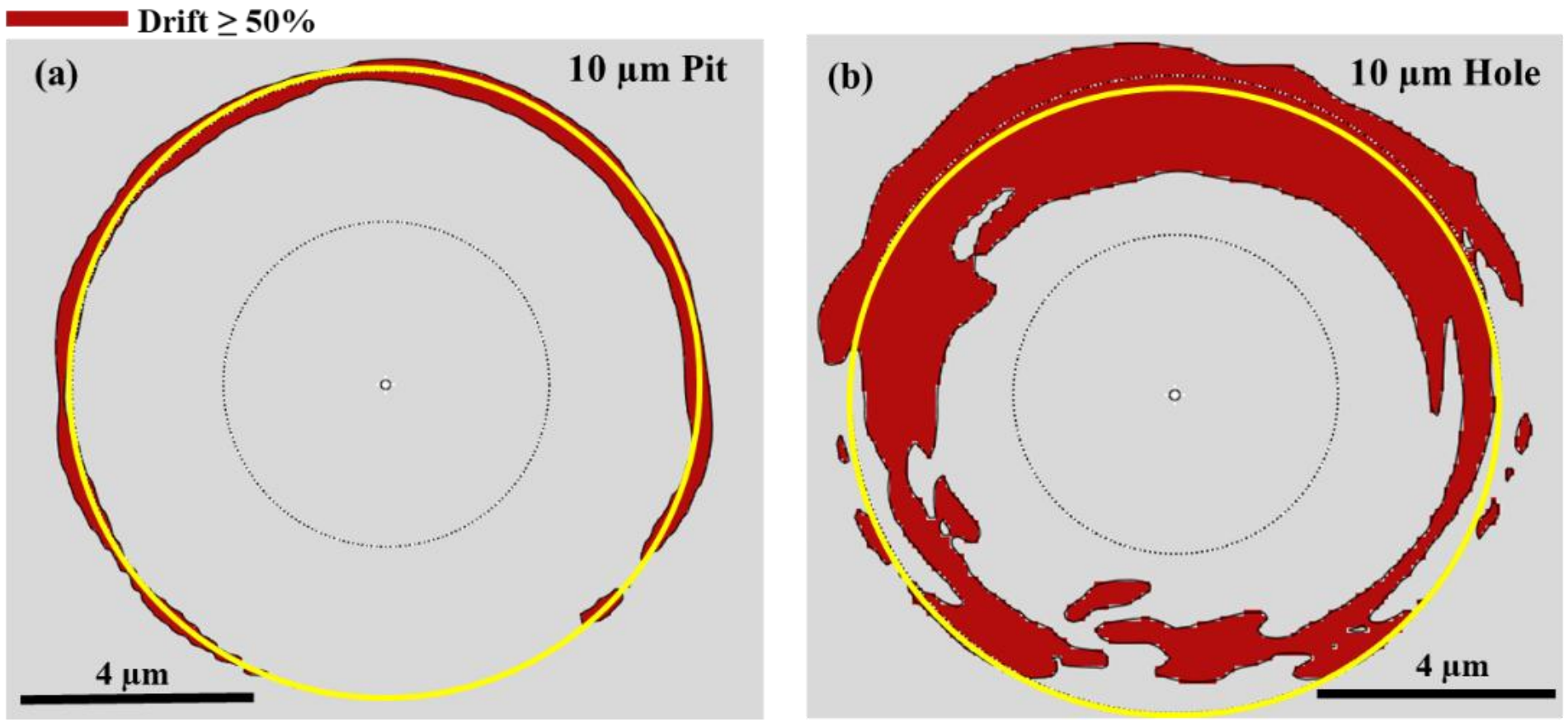

Fig. 5 **Exciton current ratio $\lambda$ maps.** (a) Pit – negligible funneling area is calculated at the circumference of the pit and is mostly suppressed by the $SiO_2$ substrate. (b) Hole – large area funneling is observed mostly in the suspended area (inside the yellow circle that depicts the circumference). The area value in the main text regards only to the area inside the hole and not on the SiN substrate.

We confirm our result with a numerical model following Ref.[9]. Details of the model are in the Methods section. We define the parameter $\lambda = J_{drift}/(J_{diff} + J_{drift})$ where $\lambda \geq 50\%$ means that the drift of the excitons $J_{drift}$ is dominant and therefore funneling is favorable over diffusion $J_{diff}$ (and exciton-to-trion conversion is suppressed). In Fig. 5 we present a binary map based on the numerical model with input from the strain map derived from the AFM data, showing the area of dominant drift ($\lambda \geq 50\%$) both for the pit and the hole. We find from the numerical model that the funneling area is $\sim 34\mu m^2$. The discrepancy between the experimental value of $\sim 42\mu m^2$ and the calculated one can be understood from the difference in obtaining the funneling area. While we derived the experimental funneling area from the PL data, the numerical model relied on the AFM data. The difference between the datasets is responsible for the apparent differences. Nevertheless, the numerical model confirms that funneling is dominant for the hole compared to the pit.

In conclusion, in this work we resolved a standing problem in the field of mechanical-strain funnels in TMDs. Is funneling dominant? Is exciton-to-trion conversion dominant? Can we achieve large area funneling? We found that with asymmetrical strain profiles that can be achieved in large-diameter through-holes this is indeed the case. This special geometry takes advantage not only of the asymmetrical strain profile that generates high SG over a large annular region but also suppresses exciton-to-trion conversion to a small region, the process that has been the limiting factor in large-area funneling. This work opens the avenue for excitonic transport in large area in suspended TMDs.

## Methods

### Exfoliation

Large area flakes necessary for this work were exfoliated with the help of an Au sacrificial layer. At first 50 $nm$ Au/$SiO_2$/Si was coated with Polyvinylpyrrolidone (PVP) solution, and a thermal release tape (TRT) was attached to it, and the Au layer is quickly peeled with the help of the TRT tape. The assembly TRT/PVP/Au was placed and gently rubbed over the shining part of the parent $WSe_2$ crystal and then removed from it. Then the final assembly TRT/PVP/Au/Flake was placed over the polydimethylsiloxane (PDMS) and heated up to $130^oC$ for the detachment of the TRT tape. In the final step of the exfoliation to clean the PVP the assembly PVP/Au/Flake/PDMS was cleaned with deionized water, and the Au was thoroughly removed by employing a gold etchant (KI/$I_2$ solution). Special care was taken to remove the PVP and Au. Finally, the flake/PDMS was inspected with a microscope.

### Transfer

The flake/PDMS assembly was used to transfer the flake from the PDMS to the desired holes and pits by a 3-axis transfer stage equipped with an Olympus optical microscope. More details about the fabrication of holes and pits are given in section I of the SI. The transfer was done at $25-40^oC$. SiN suspended membranes are more delicate than $SiO_2$/Si substrate. Both the touchdown of the PDMS over the SiN membrane and lift off was done very slowly, the entire transfer process takes about 45 minutes.

### AFM and SEM measurements

After the completed transfer, AFM measurements were done by using a $\sim 10\ nm$ radius Si tip in tapping mode to ensure that the interaction between the tip and the suspended layer is minimal to avoid destruction of the suspended monolayer. Large $\sim 10\ \mu m$ suspended holes were especially suspectable to damages. A commercial Bruker AFM was used for both imaging and for obtaining force displacement curves (for extracting the adhesion properties – see SI). SEM imaging was performed at low accelerating voltage of $1-2\ kV$ to minimize charging and was performed only after optical measurements.

Following Eq. 3 in the main text, we extract the strain at each AFM measurement pixel by:

$$\varepsilon = \sqrt{1+\left(\frac{\partial z}{\partial x}\right)^2+\left(\frac{\partial z}{\partial y}\right)^2}-1 \quad (4)$$

The slope $\frac{\partial z}{\partial x}$ and $\frac{\partial z}{\partial y}$ was computed from the AFM height profiles and was used to get the strain values by employing Eq. 4[28,37]. Once the strain map is computed, SG map is generated by employing the gradient operator ($\boldsymbol{\nabla}$) in radial coordinates. The SG map is computed along the radial direction ($\frac{\partial \varepsilon}{\partial r}\hat{\boldsymbol{r}}$ ) for both the pit and the hole while maintaining the origin at the center of the pit/hole.

### PL and Raman

PL and Raman mapping, single spectra were recorded with a Horiba Labram system. 1800 lines/mm and 600 lines/mm gratings were used in these experiments, with a HeNe 633 $nm$ excitation. Mapping duration was kept below 3-4 hours to reduce thermal and mechanical instabilities of the measurement setup. The 520 $cm^{-1}$ Si Raman peak was used to calibrate the spectrometer each time before any measurements. A 100x objective with 0.9 NA was used for better collection efficiency, giving a laser spot size of $\sim 700\ nm$. The optical mapping was done with either 250 or 500 nm step size (less than the laser spot size). For power variations the laser power was varied from 1.2 $\mu W$ to 1200 $\mu W$.

The strain map derived from the exciton emission energy was computed by employing a gauge factor of $G = 49\ meV/\%$ shift for uniaxial strain in $WSe_2$[38]. $E_0$ (the exciton energy at zero strain) was taken from the spectra from the center of the hole/pit. Similarly, the strain map from the 2LA(M) map was computed by assigning $\omega_0 = 262\ cm^{-1}$ position at zero strain from the center of the hole. The shift from $\omega_0$ along with a gauge factor of $0.95\ cm^{-1}/\%$ was employed[36]. 2LA(M) was chosen as it is very sensitive to strain and has relatively high intensity than other Raman peaks. The SG map from the PL and Raman was generated by taking into account the stage movement steps employed for the maps.

The strain in ij[th] pixel computed from the exciton map is given by:

$$\varepsilon_{ij} = \frac{E_0 - E_{ij}}{G} \tag{5}$$

and the SG is given by in x-y coordinate system:

$$|\nabla\varepsilon| = \sqrt{\left(\frac{\partial \varepsilon}{\partial x}\right)^2 + \left(\frac{\partial \varepsilon}{\partial y}\right)^2} \tag{6}$$

Before computing the SG, the strain values were smoothened, as numerical differentiation can introduce noise. Similar approach is employed for strain and SG from Raman mode. The strain map in x-y or SG map in x-y is transformed to polar coordinates by employing these two steps - 1) an appropriate Jacobian and 2) by incorporating the optical image of the hole/ pit to identify the center of the hole/pit. As the radial map depends upon the center of the hole/pit this step is crucial to avoid any artifacts.

Numerical model

The continuity equation for excitons (behaving as currents) is given by:

$$\frac{\partial n}{\partial t} = -\nabla \cdot \mathbf{J} + S - \frac{n}{\tau} - R_A n^2 \tag{7}$$

Where $n$ is excitons per unit area, $\boldsymbol{J}$ is total exciton current, $S$ is the exciton generation rate due to laser excitation ($\mu m^{-2}\, ns^{-1}$) and $R_A$ is the Auger coefficient. The exciton current is given by:

$$\mathbf{J} = \mathbf{J}_{\mathrm{diff}} + \mathbf{J}_{\mathrm{drift}} = -D\,\boldsymbol{\nabla} n - \mu\, n\, \boldsymbol{\nabla} u, \qquad \mu = \frac{D}{k_B T} \tag{8}$$

$\boldsymbol{J}_{diff}$ is the exciton current due to random exciton diffusion and $\boldsymbol{J}_{drift}$ is exciton drifting due to the potential generated by the strain gradient. $D$ is diffusion coefficient, $\mu$ is mobility of the excitons and -$\boldsymbol{\nabla} u$ is the force acting on the exciton and $u = E_g$ the excitonic band gap energy. Thus, **F** can be expressed as Eq. 2. As our PL maps measured by doing a raster scanning the numerical model also uses a raster scanning model for excitation. At steady state when $\frac{\partial n}{\partial t} = 0$ Eq. 7 in cartesian coordinates becomes:

$$\frac{\partial}{\partial x}\left[D\frac{\partial n}{\partial x} + \mu\, n\frac{\partial u}{\partial x}\right] + \frac{\partial}{\partial y}\left[D\frac{\partial n}{\partial y} + \mu\, n\frac{\partial u}{\partial y}\right] - \frac{n}{\tau} - R_A n^2 + S(x,y) = 0 \tag{9}$$

The values used to solve the steady state equation: $D = 0.3\frac{cm^2}{s}$, $\tau = 1.1\, ns$, $\mu = \frac{D}{K_B T} = 12\frac{cm^2}{Vs}$, $n_0 = 10^8\ cm^{-2}$ (not applicable when Auger is neglected). FWHM of the laser spot is 0.7 $\mu m$, and the grid size is $\Delta x, \Delta y = 88 nm$. Also, for simplicity we employed cartesian coordinates for calculation and neglected the Auger term. For a continuous wave laser, the exciton generation is linked with S which is:

$$S(x,y) = \exp\left[-\frac{(x-x_s)^2 + (y-y_s)^2}{2\sigma^2}\right], \quad \sigma = \frac{\mathrm{FWHM}}{2\sqrt{2\ln 2}} \tag{10}$$

In raster scanning (as employed in PL map) the SG of each pixel is taken from the AFM data and $n$ and $\nabla n$ of is computed for each pixel from Eqs. 9-10. Finally, Eq. 8 is used to get the excitonic currents – drift and diffusion and the ratio between them - $\lambda = \boldsymbol{J}_{drift}/(\boldsymbol{J}_{diff} + \boldsymbol{J}_{drift})$.

**Custom code and analysis**

The code used for analysis and computation of the strain, SG, from the AFM data, and from the PL and Raman data were prepared by using Claude, Opus 5.5 model and the code was run in MATLAB. Output of the code behind each analysis was thoroughly checked and cited wherever needed.

**Data availability**

Data presented in the article and in SI is available from the corresponding author upon reasonable request.

**Code availability**

The custom codes employed to produce the results in this paper are available from the corresponding authors upon reasonable request.

**Acknowledgements**

The authors would like to acknowledge the help of Dr. Nitzan Maman for perforating the holes in the SiN with the FIB. M. G. H. would like to acknowledge funding from the ISF grant no. 492/25, from the Israel Ministry of Energy (MOE) 223-11-059 and BSF project 2024124.

**Author contributions**

N. B. and M. G. H. conceived the project, performed the data analysis and wrote the paper. S. K. and M. P. fabricated the samples. S. K. and N.B performed the AFM scans. N. B. conducted the optical experiments. M. G. H. provided funding and supervised the project. All authors revised the paper.

**Competing interests**

The authors declare no competing interests.

Supplementary Information

**Large area exciton funneling in suspended monolayer $WSe_2$**

*Nilanjan Basu [1], Svyatoslav Kostyukovets [1], Mousri Paul [1], Moshe G. Harats *[1,2]*

*[1]Department of Materials Engineering, Ben-Gurion University of the Negev, Be'er Sheva 8410501, Israel*

*[2]Ilse Katz Institute for Nanoscale Science and Technology, Ben-Gurion University of the Negev, Beer-Sheva 8410501, Israel*

** E-mail: mharats@bgu.ac.il*

# Contents

## I. Exfoliation and hole/pit fabrication

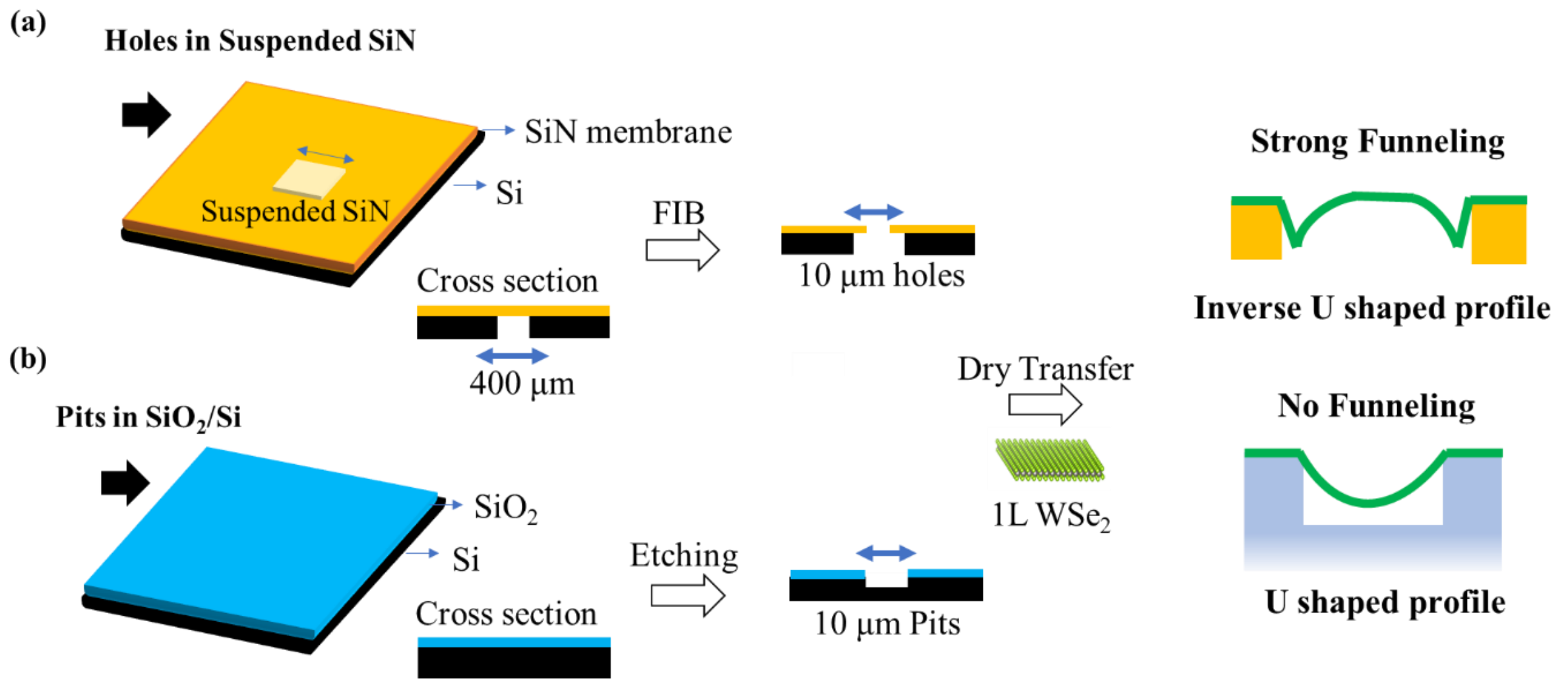

**Fig. S1 Suspended 1L $WSe_2$ transfer over holes in SiN membrane and pits in $SiO_2$/Si.** (a) Stepwise fabrication of holes in suspended SiN membrane and (b) pits in $SiO_2$/Si substrate.

Fabrication of holes and pits: Fig. S1a shows the schematic of making holes in SiN membrane. The Suspended SiN membrane is $400 \times 400\mu m^2$ with $50nm$ thickness. Focused Ion Beam (FIB) was used to etch holes of different diameter ($3{,}5{,}10\mu m$) at the center of the squared shaped suspended SiN membrane.

Fig. S1b shows the schematic of making pits in $285nm$ $SiO_2$/Si substrate. The depths of the pits were 1.5 µm. The pits array layouts were designed in KLayout prior to lithography. A 5 × 5 mm silicon chip with a 285 nm thermal $SiO_2$ layer was cleaned sequentially with acetone and isopropyl alcohol (IPA), dried under a dry nitrogen , and spin-coated with AZ 1505 positive photoresist at 4000 rpm (400 rpm/s acceleration) for 50 s, followed by a prebake at 100°C for 90 s. Pit patterns were exposed using a maskless aligner (Heidelberg Instruments MLA 150), developed in AZ 726 MIF for 30 s, rinsed with deionized (DI) water, dried with nitrogen, and hard-baked at 95°C for 5 min. Following a 1 min $O_2$ plasma descum, the exposed 285 nm $SiO_2$ layer was etched for 45 seconds by inductively coupled plasma reactive ion etching (ICP-RIE; Oxford Instruments PlasmaPro 100 Cobra) using a octafluorocyclobutane/acetylene ($C_4F_8/C_2H_2$) gas mixture. To achieve the desired pit depth, the underlying silicon was subsequently etched for 45 s using a deep reactive ion etching system (Oxford Instruments PlasmaPro 100 Estrelas) with a sulfur hexafluoride/octafluorocyclobutane ($SF_6/C_4F_8$) gas mixture. Residual photoresist and fluorocarbon polymers were stripped in piranha solution ($H_2SO_4/H_2O_2$, 3:1 volume ratio) using a semiautomatic wet-processing platform (SUSS MicroTec HMxSquare), after which the substrates were thoroughly rinsed with DI water, acetone, and IPA, dried under nitrogen, and treated with $O_2$ plasma for 5 min prior to transfer.

Exfoliation and transfer of 1L $WSe_2$: Monolayers of $WSe_2$ were exfoliated using gold-assisted exfoliation over PDMS[1]. Standard dry transfer technique was employed to transfer the flake over desired substrates after thoroughly cleaning the Au layer with Au etchant. Transfer was done at 40°C. A 150 nm gold (Au) film was deposited over a polished Si wafer, followed by spin coating a 10 wt% solution of polyvinylpyrrolidone (PVP K-30, SRL) in a 1:1 weight-ratio acetonitrile/ethanol mixture at 3000 rpm (1000 rpm/s acceleration) for 1 min and baking on a hot plate at 150°C for 1 min. Thermal release tape (TRT; Semiconductor Equipment Corp., RAY-4LSC(N)) was applied over the PVP/Au stack and lifted to form a smooth gold tape, which was stored until use and peeled just prior to exfoliation to minimize air exposure of the freshly exposed gold surface. The smooth gold tape was pressed against a freshly cleaved $WSe_2$ crystal and heated at 110°C for 20 min to release the first TRT layer. A second TRT strip was applied to flatten the Au layer on the exfoliated transition-metal dichalcogenide (TMD) flake before lifting and transferring the assembly onto a PDMS stamp supported on a glass slide. The stamp consisted of a commercial Gel-Pak film (PF-40x40-0065-X4) mounted atop a custom adhesive PDMS base prepared from a Sylgard 184 elastomer kit (40:3 base-to-curing-agent ratio). Heating at 110°C for 5 min released the second TRT layer, after which the sample was immersed in deionized water for at least 3 hours (with water exchanged every 30 min) to dissolve the PVP layer. The underlying gold film was then removed in a commercial gold etchant ($KI/I_2$) at room temperature, rinsed sequentially with distilled water and isopropyl alcohol (IPA), dried with dry nitrogen, and dry-transferred onto the target substrate.

## II. SEM and AFM images of hole/pit

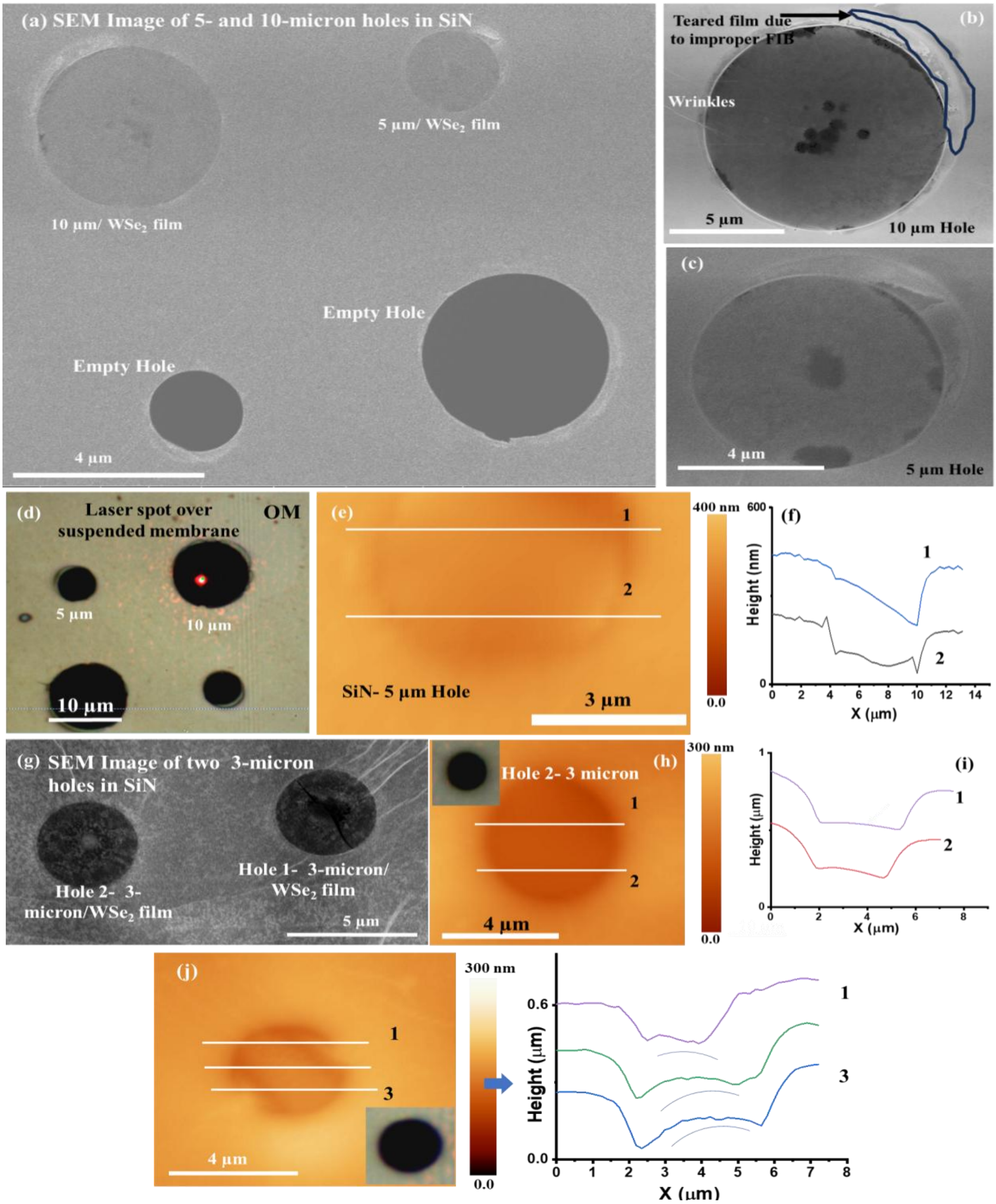

**Fig. S2 SEM and AFM images of monolayer $WSe_2$ over holes in SiN membrane.** (a), (b), (c) SEM image and (d) optical micrograph of the suspended flake over $10\mu m$ and $5\mu m$ holes. (e), (f) AFM image and height profile of $5\mu m$ hole. (g) SEM image of the suspended flake over two different $3\mu m$ holes. (h), (i) and (j) AFM image and height profiles of the corresponding $3\mu m$ holes.

A single large flake was transferred over the 10 and $5\mu m$ holes (a similar large flake was transferred over $3\mu m$ holes). In the SEM image in Fig. S2a,b wrinkles can be seen over the $10\mu m$ hole and top right part of the hole had non uniform coverage of the flake due to improper FIB. The SEM images show that the flakes are continuous and have a uniform coverage. The optical micrograph in Fig. S2d shows the laser spot over the suspended flake. The AFM image and the corresponding height profile of the $5\mu m$ hole in Fig. S2e,f shows that the flake has a U-shaped profile.

The SEM image of the $3\mu m$ holes as shown in Fig. S2g shows a rupture in hole 1 and intact flake in hole 2. The AFM images and corresponding height profiles of the holes are shown in Fig. S2h,j. The rupture is not seen in the adjacent AFM image and appears to occur during SEM imaging after the PL, Raman and other measurements were completed. All the SEM images were recorded after all optical and AFM measurements so that the charging does not affect the measurements. The AFM height profiles of the $3\mu m$ holes as shown in Fig. S2i,j also shows a U-shaped profile. Although U-shaped, a upward curvature can be seen which is due to the air pressure.

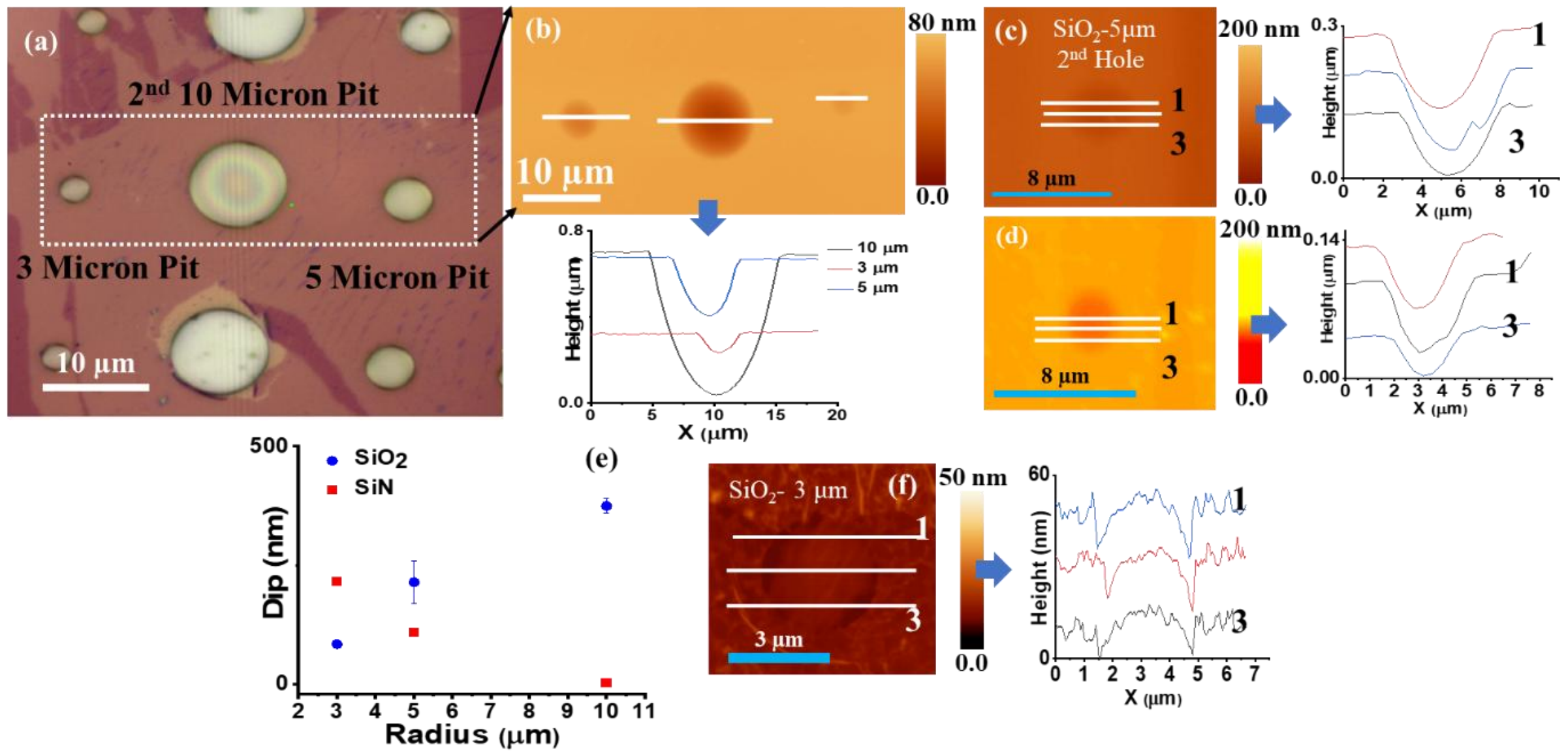

**Fig. S3 Optical micrographs and AFM images of pits in SiO2/Si.** (a), (b) Optical micrograph of a 2nd $10\mu m$ pit, $3\mu m$ and $5\mu m$ pits, and corresponding AFM image and height profile. (c), (d) AFM image of 2nd set of $3\mu m$ and $5\mu m$ pits. (e) Comparison of the of depth of sag from the circumference for both pits and holes. (f) AFM image and corresponding height profile of a ruptured flake over a $3\mu m$ pit.

The U-shaped sag is repeated in pits of dimension 3,5, and $10\mu m$ as shown in Fig. S3a-d. For pits it shows a linear relationship and for holes inverted U shaped profile can be seen only for a $10\mu m$ hole. The comparison is shown in Fig. S3e. Inverse U shaped profile is also seen in a $3\mu m$ pit when the flake is ruptured at two sides as shown in Fig. S3f. This is because of air pressure equalization through the rupture.

## III. Force displacement curves of SiN and $SiO_2$

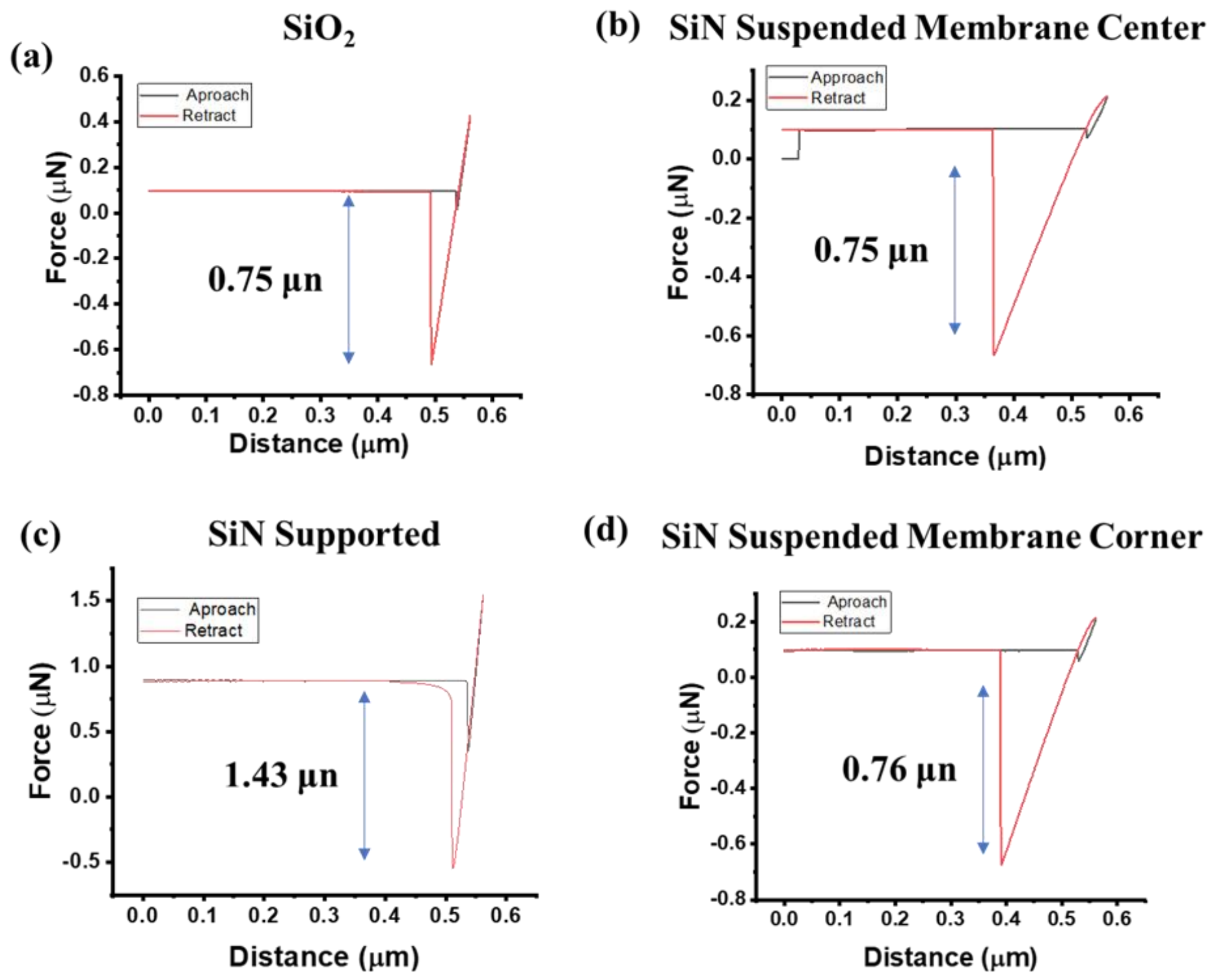

**Fig. S4 Force displacement curves**. (a) Over $SiO_2$ /Si, over (b) suspended SiN membrane center, (c) SiN/ Si and (d) SiN suspended membrane center.

The force displacement curve of a surface obtained from AFM tip interaction with the surface gives a glimpse of the adhesion or pull force of that surface[2]. Fig. S4 shows the force displacement curve of different surfaces. The AFM tip radius employed for nanoindentation in $50nm$. The adhesion of $285nm$ thick $SiO_2$/Si is $75\mu N$. For the supported SiN, that is SiN over Si shows the highest adhesion of $143\mu N$. It becomes nearly half when measured over suspended SiN membrane (both at the center and at the corner of the suspended membrane). This is expected as suspended SiN is free from any underlying substrate and can act freely with the AFM tip.

The adhesion of the suspended SiN and $SiO_2$ is similar. This shows that the $WSe_2$ flake architecture and shape over the pits and holes is independent of the adhesion force between the underside cylindrical (well area of the pit and hole) area.

## IV. Asymmetric and uniaxial strain.

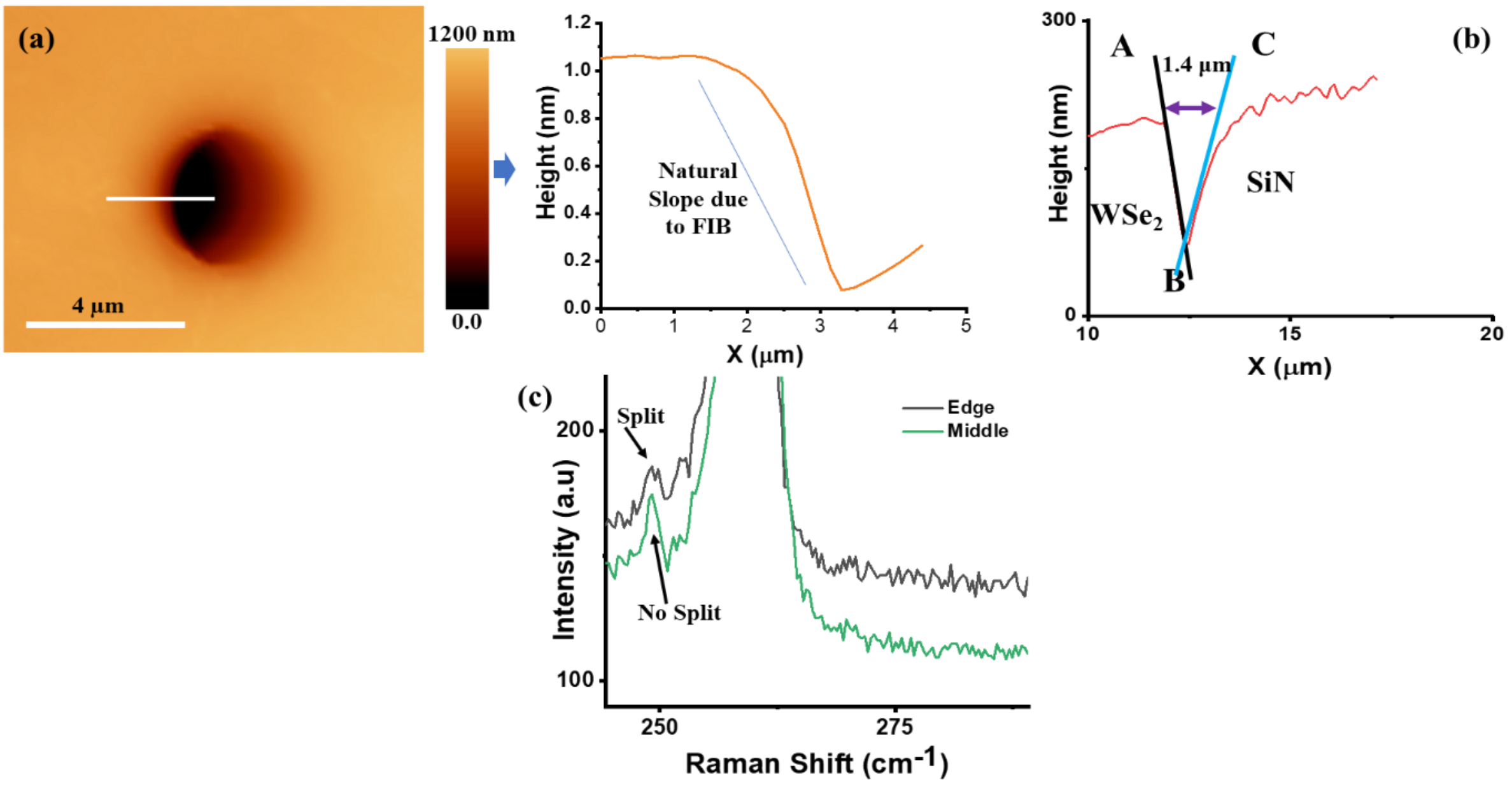

**Fig. S5 Empty hole and uniaxial strain.** (a) AFM image of an empty hole with corresponding height profile. (b) A typical AFM height profile of monolayer $WSe_2$ on a $10\mu m$ hole in SiN showing the SiN side and the $WSe_2$ side. (c) Raman split of the E peak around $250cm^{-1}$ under uniaxial strain

Holes made from FIB naturally have an intrinsic slope of the wall as shown in Fig. S5a for an AFM measurement of an empty hole. When measuring a hole that is covered with a monolayer, two different slopes on both sides (one from the hole surface and other from the rise of the flake) gives rise to the asymmetry as seen in Fig. S5b. Fig. S5c shows the in-plane Raman peak (E peak) near $250cm^{-1}$. The E peak exhibits a split under uniaxial strain[3].

## V. Further strain maps

Fig. S6a show the denoised height profile, corresponding geometric strain (calculated from the Green-Lagrange formula) and strain gradient in $x$ direction for ½ section of the $10\mu m$ pit. The height profile was denoised before computing strain and strain gradient because $(dz/dx)2$ can abruptly magnify local noises in the AFM data. The strain maximizes about $0.8\mu m$ from the edge but the SG maximizes earlier. As we move toward the pit center the geometric strain approaches zero and very low SG. Fig. S6b shows similar analysis of the ½ section of the $10\mu m$ hole. The height profile shows the trench after the circumference of the hole. The strain profile shows that as we move from the circumference towards the hole center the strain initially increases to a maximum of 1.6% and becomes 0% at the bottom of the trench and then it increases when we move away from the bottom and then decreases as we move towards the hole center. Interestingly the SG also follows a similar pattern like strain, but it has a phase difference. Which means that areas of maximum strain do not necessarily give rise to maximum SG and vice versa. Beginning of the trench, high SG area has a lateral span of $\sim 2.4\mu m$, after this the SG vanishes as we approach the center of the hole. This is also evident in the PL map of the $10\mu m$ hole, where high degree of funneling is observed where high SG is recorded. Fig. S6c-e shows the strain map (computed by employing green-LaGrange equation for geometric strain in cartesian coordinates) of 3,5 and $10\mu m$ pits. The strain maps show that there is a homogeneous distribution of the strain, meaning negligible SG and negligible strain at the center of the pit. Fig. S6f shows the strain map of $3\mu m$ holes where strain is maximum at the circumference of the holes, indicating a localized SG which is insufficient for funneling over large area.

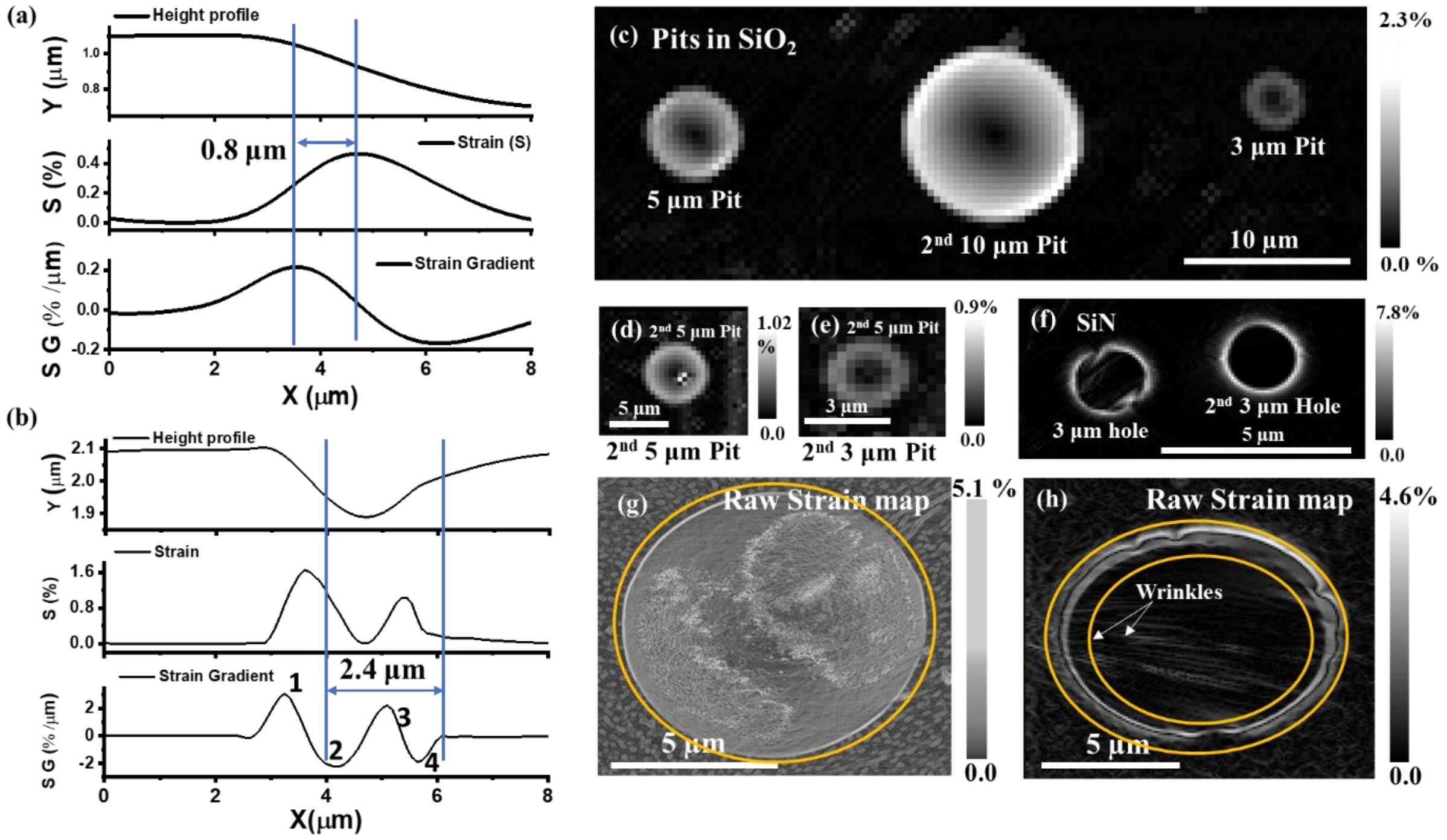

**Fig. S6 Strain profile and strain maps deduced from AFM measurements.** (a, b) Height profile, strain and SG line cut of $10\mu m$ pit and hole, respectively. (c-e) Strain map of different size pits. (f) Strain maps of $3\mu m$ holes. (g-h) Raw strain map (without denoising AFM line profiles) of the $10\mu m$ pit and hole, respectively.

Fig. S6g,h shows the strain map of the $10\mu m$ pit and hole from raw AFM data without denoising. As the AFM data is not smoothened out local variations abruptly magnify the strain map, which affects the interpretation. To properly interpret the strain and SG map the raw AFM data must be smoothened out so that local noises in line profile are minimized and the original nature of the topography is dominant. A Savitzky-Golay filter was employed for filtering in the horizontal direction and a Gaussian filter was employed in the lateral direction. Once the local noises in each line profile were removed the strain and SG maps were computed.

## VI. PL spectra under different laser power and further PL maps

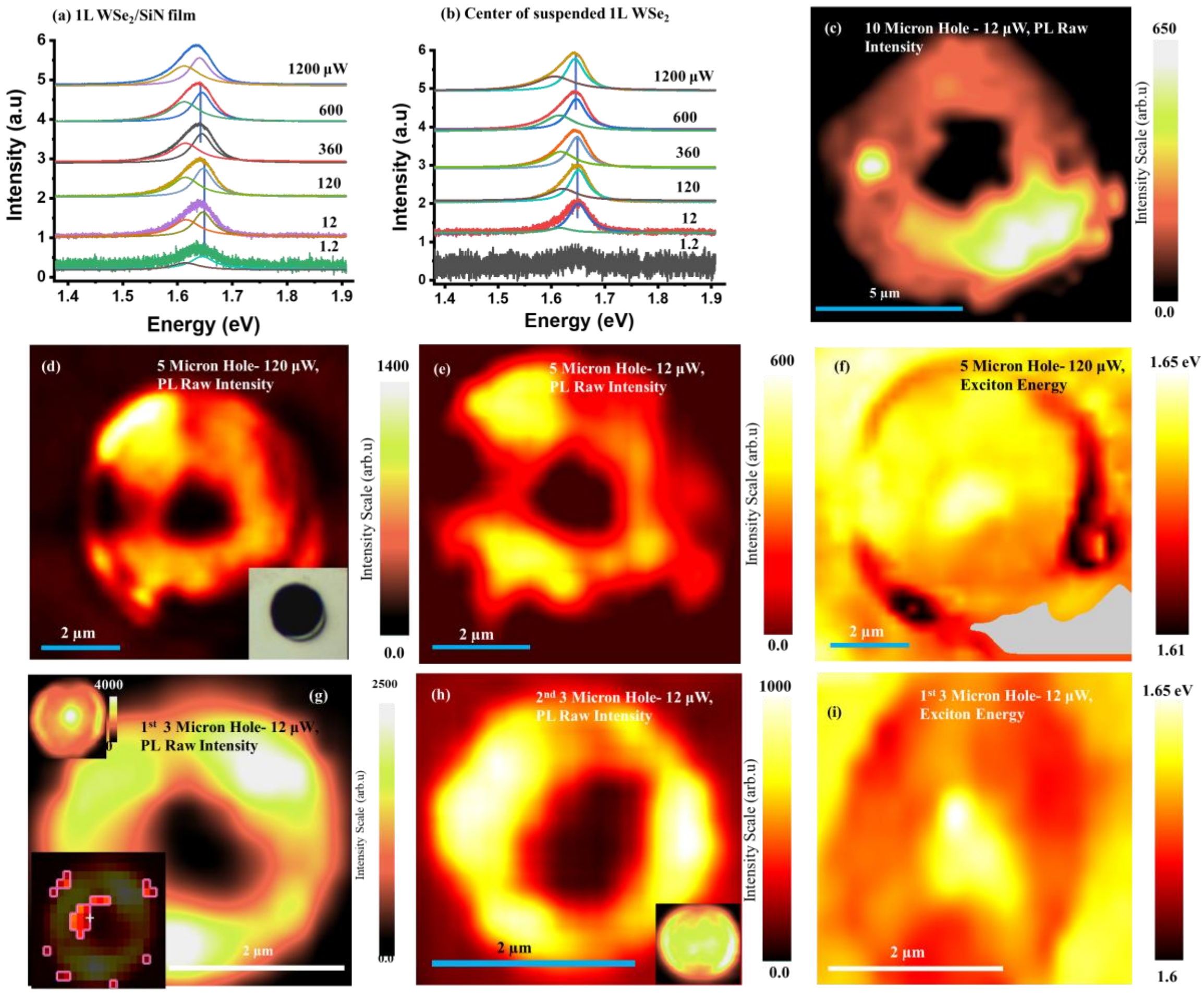

**Fig. S7 PL spectra under different laser power of flakes over holes.** (a), (b) Normalized PL spectra shifted for clarity at different powers from a flake over SiN (substrate) and from the center of the suspended flake, respectively. Powers are in $\mu W$. (c) Area scan of the maximum PL intensity of the suspend flake over $10\mu m$ hole at $12\mu W$ laser power. (d-f) Area scan of the maximum PL intensity of a suspended flake over a $5\mu m$ hole at $120\mu W$ (d) and $12\mu W$ (e) laser power and exciton energy area scan at 120 μW laser power (f). (g) Area scan of the maximum PL intensity of a suspended flake over a $3\mu m$ hole, inset shows the area scan at $120\mu W$, and (h) corresponding exciton energy area scan at $12\mu W$. (i) Area scan of the maximum PL intensity map of the flake over an additional $3\mu m$ hole at $12\mu W$, inset shows the map at $120\mu W$.

Fig. S7a, b shows the PL spectra as a function of excitation power of the monolayer WSe2 supported on SiN and from the center of the suspended flake over the $10\mu m$ hole. We fit the spectra to the exciton line and the trion line with the electron-recoil effect[4,5] to resolve the evolution of the emission as a function of excitation power. At both measurement spots up to $120\mu W$ there is almost no shift of the exciton and trion peak. From $360\mu W$ a red shift is observed, which is due to thermal effects. Therefore, our measurements in the main manuscript were limited to $120\mu W$. Fig. S7c shows the PL intensity map of the suspended flake over a $10\mu m$ hole at $12\mu W$. The intensity map is similar to that of $120\mu W$ presented in the main manuscript. PL intensity map of the suspended flake over a $5\mu m$ hole at $120\mu W$ and $12\mu W$ is shown in Fig. S7d-e, respectively, the inset shows the optical micrograph. The exciton energy map is shown in Fig. S7f. The 5 μm hole also behaves in similar fashion regarding PL intensity distribution (annular ring pattern). A circular pattern at the hole circumference can be observed in the exciton energy map as the exciton energy shifts due to strain. But no considerable and uniform exciton funneling is observed. This is due to lack of high SG as the AFM profile shows a flat profile and not an inversed U-shape (see Fig. S2f).

Fig. S7g,h presents the PL intensity map of two different $3\mu m$ holes at $12\mu W$, the inset shows the map at $120\mu W$ laser power. At $12\mu W$ the PL map of the $3\mu m$ hole behaves in a similar manner as the 5 and $10\mu m$ holes, but at $120\mu W$ the intensity behavior is reversed (it shows uniform brightness or even brighter emission at the center). The laser spot has a diameter of $700nm$ and the exciton diffusion length at RT for a suspended flake is about $\sim 0.4\mu m$, which means that the excitation spot is covering more than 1/3 of the hole diameter. At higher power, exciton diffusion is enhanced due to exciton-exciton repulsion[6] which in return makes the center part brighter as there is negligible strain. Fig. S7i shows the exciton energy map of the 1st $3\mu m$ hole (Fig. S7g) at $12\mu W$. As expected, the center section shows no energy shift and a uniform shift is observed in the circular annular part. As in the case of the $5\mu m$ hole, the $3\mu m$ holes did not show considerable funneling due to a lack of high SG as evident from the AFM profiles presented in Fig. S2j.

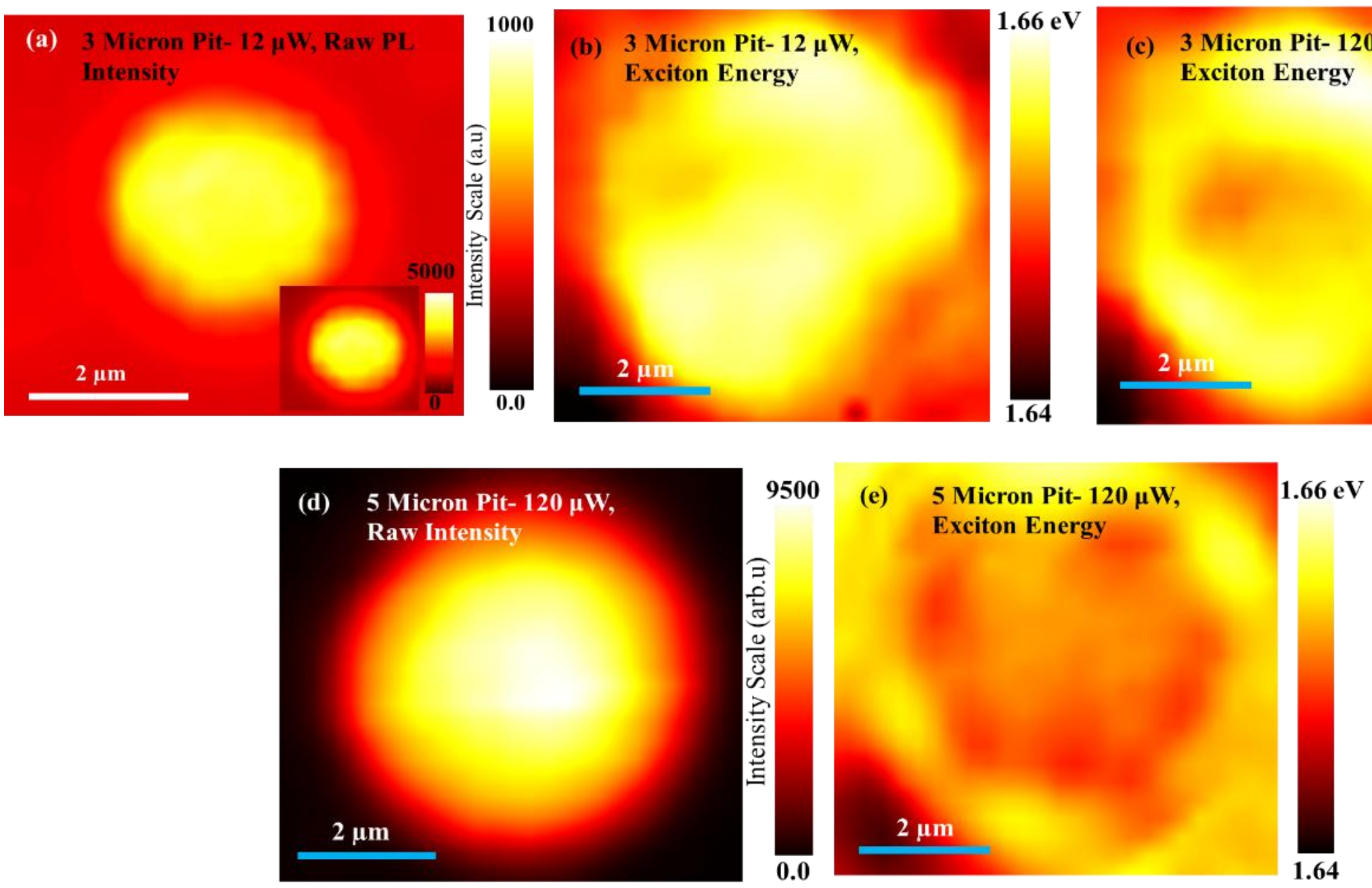

**Fig. S8 PL spectra of flakes over pits of 3 and $5\mu m$.** (a) Area scan of the PL intensity of a flake over a $3\mu m$ pit at $12\mu W$, inset shows same measurement $120\mu W$. (b, c) Exciton energy map of the 3 µm pit at $12\mu W$ and $120\mu W$, respectively. (d, e) Area scan of the PL intensity of a flake over a $5\mu m$ pit at $120\mu W$ and corresponding exciton energy map.

Fig. S8a shows the PL map of a $3\mu m$ pit at $12\mu W$ and at $120\mu W$. The PL map of the pit at both high and low power behaves similarly and does not change dramatically from the results of the $10\mu m$ pit presented in the main manuscript. Fig. S8b,c shows the exciton energy map of the $3\mu m$ pit at two different powers. We see a homogeneous energy (strain) distribution with the center having slightly low energy as there is low strain at the center of the pit. Fig. S8d,e shows the PL intensity and exciton energy map at $120\mu W$, respectively. The PL map shows a similar homogeneous intensity distribution as in the $3\mu m$ and $10\mu m$ pits. The exciton energy map in Figure S9e shows a peak shift just beside the circumference where there is high strain from the wall of the pit, but no funneling is observed as there is no dominant SG.

## VII. Further Raman measurements and comparison of SG

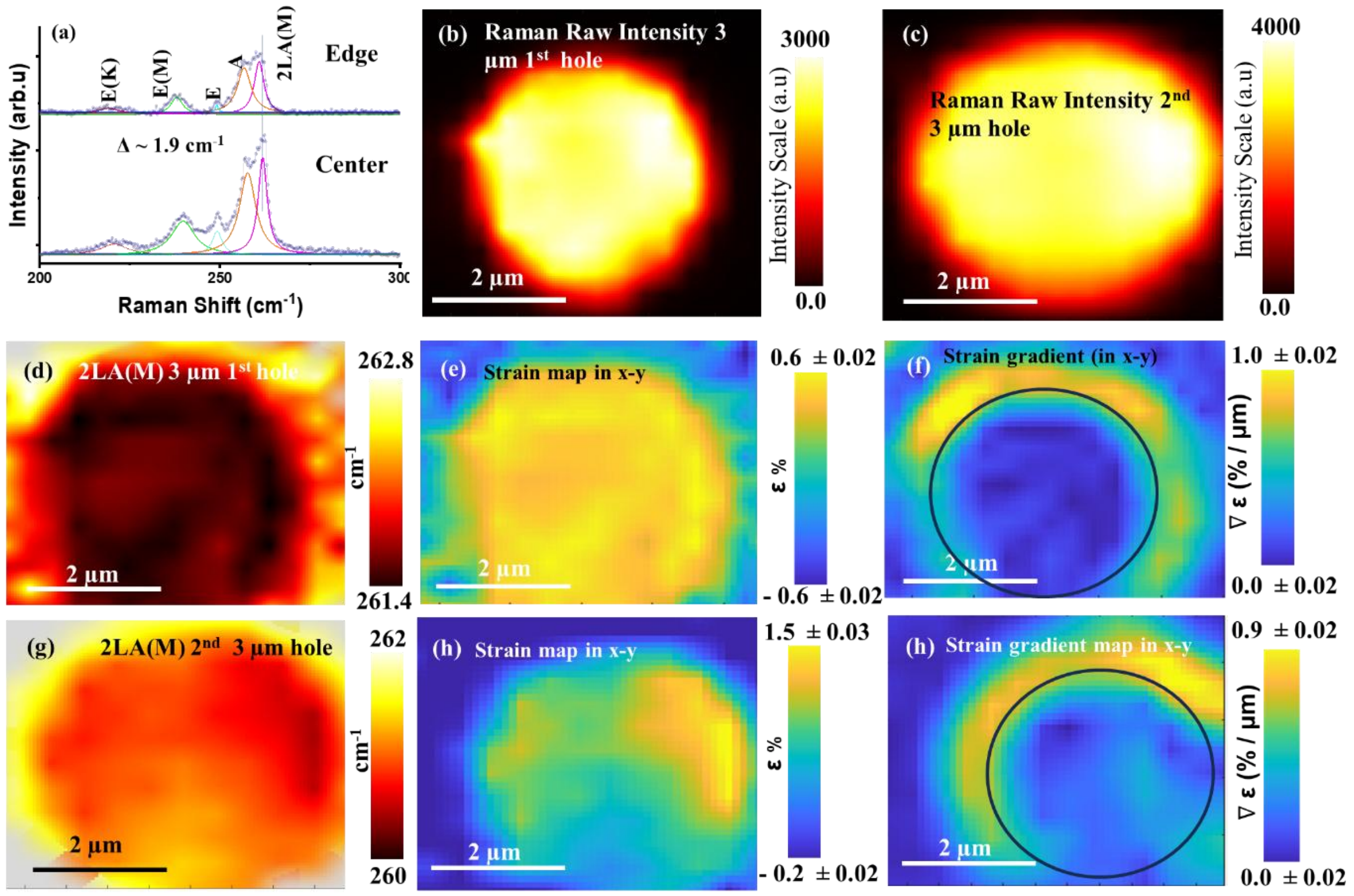

**Fig. S9 Raman peak profiles and measurement**. (a) Raman profile at the edge and at the center of the flake over $10\mu m$ hole. (b), (c) Raman raw intensity of the 1st and 2nd $3\mu m$ hole. (d-f), (g-h) 2LA (M) map and corresponding strain and strain gradient map of the 1st hole and 2nd hole, respectively. The black circle corresponds to the hole circumference.

Fig. S9a shows the Raman spectra at the center and at the edge of a $10\mu m$ hole. The major observed peaks are $E(K), E(M), E, A$ and 2LA(M) mode [7]. The 2LA(M) mode shows a shift of $1.9 cm^{-1}$ corresponding 1.6% strain. Fig. S9b,c shows the Raman intensity map of the $3\mu m$ holes, presenting a homogeneous intensity distribution. Fig. S9d shows the 2LA(M) map of the 1st 3 µm hole, and the corresponding strain map and SG map in cartesian coordinate is shown in Fig. S9e,f respectively. A homogeneous distribution is observed for the 2LA(M) mode meaning homogeneous strain in the strain map. The negative in the strain color map legend is due to the peak shift over SiN substrate. The strain gradient is also uniform and shows very low value, insufficient for funneling. Similar features are observed in the 2nd $3\mu m$ hole, as shown in Fig. S9g-i.

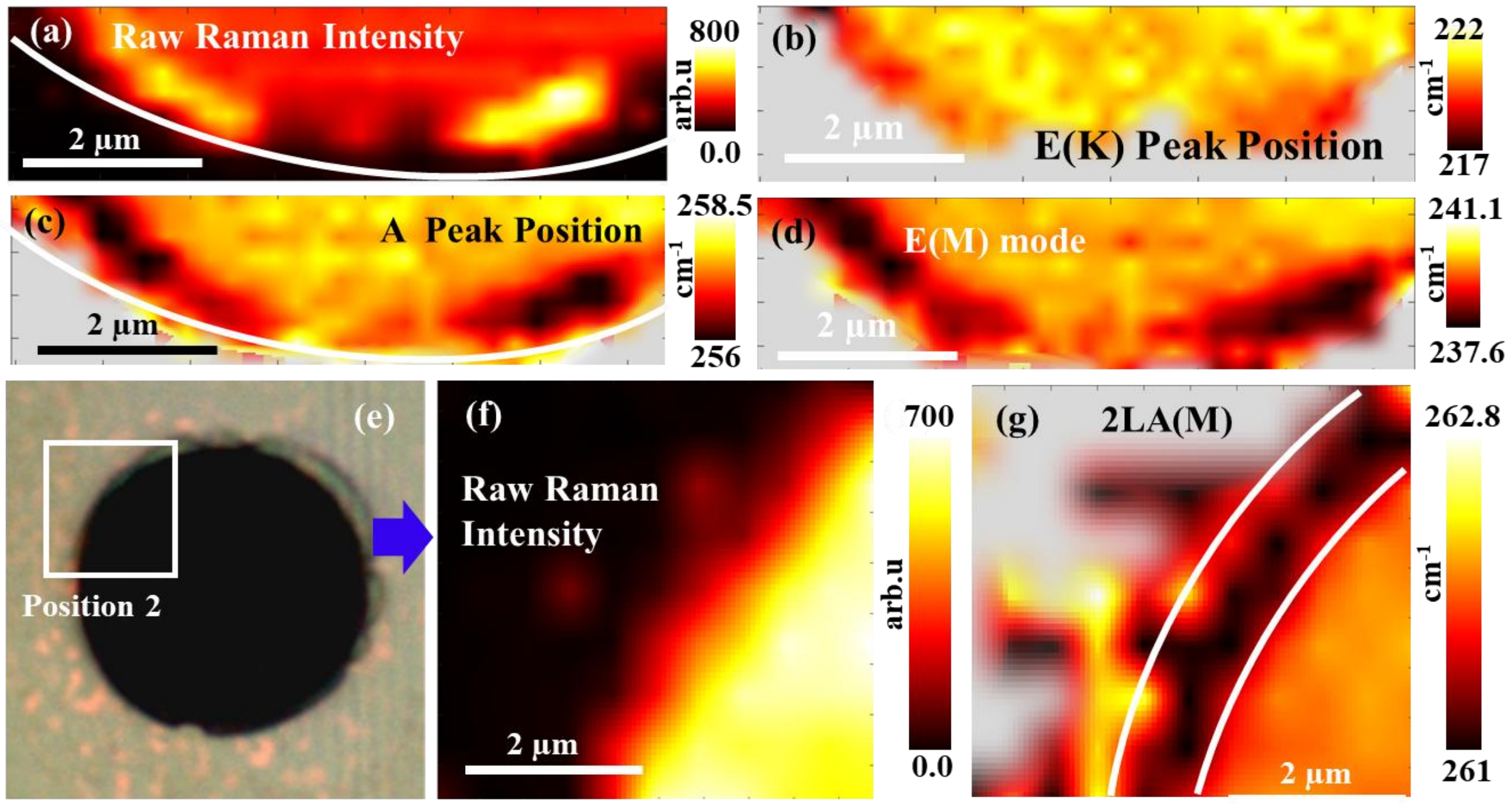

**Fig. S10 Raman measurements of a flake suspended over a 10 μm hole.** (a-d) Raman map of the lower part of the flake over $10\mu m$ hole (same area as shown in the main manuscript) showing the evolution of the $E(K), A$ and $E(M)$ modes, respectively. The white curve in (a),(c) presents the circumference. (e-g) Optical micrograph, Raman intensity and 2LA(M) peak shift map of the top left of the hole, respectively. The white curves in (g) present the area of high strain which show exciton-to-trion conversion in the main manuscript.

Fig. S10a show the Raman intensity map of the lower part of the $10\mu m$ hole (same area as shown in the main manuscript but for different modes). Fig. S10b-d shows the Raman map of $E(K), A$ and $E(M)$ modes, respectively. A clear shift in peak position is observed in all three modes. Although the strain is planer and uniaxial, the shift is observed also for the $A$ Raman mode. This is because the uniaxial strain alters the Se-W-Se bond angle and also alters the charge distribution surrounding them. As a result, it changes the lattice in horizontal direction which forces the lattice to make adjustment in the vertical direction (Poisson effect) and the A mode shifts[8]. Fig. S10e-g shows optical micrograph, Raman intensity and 2LA(M) map respectively from top left quadrant of the hole. Here a similar peak shift is observed for the 2LA(M) mode as presented in the area scan in the main manuscript. The high strain is visible in Fig. S10g between the white curves and corresponds to the exciton-to-trion conversion region in the main manuscript.

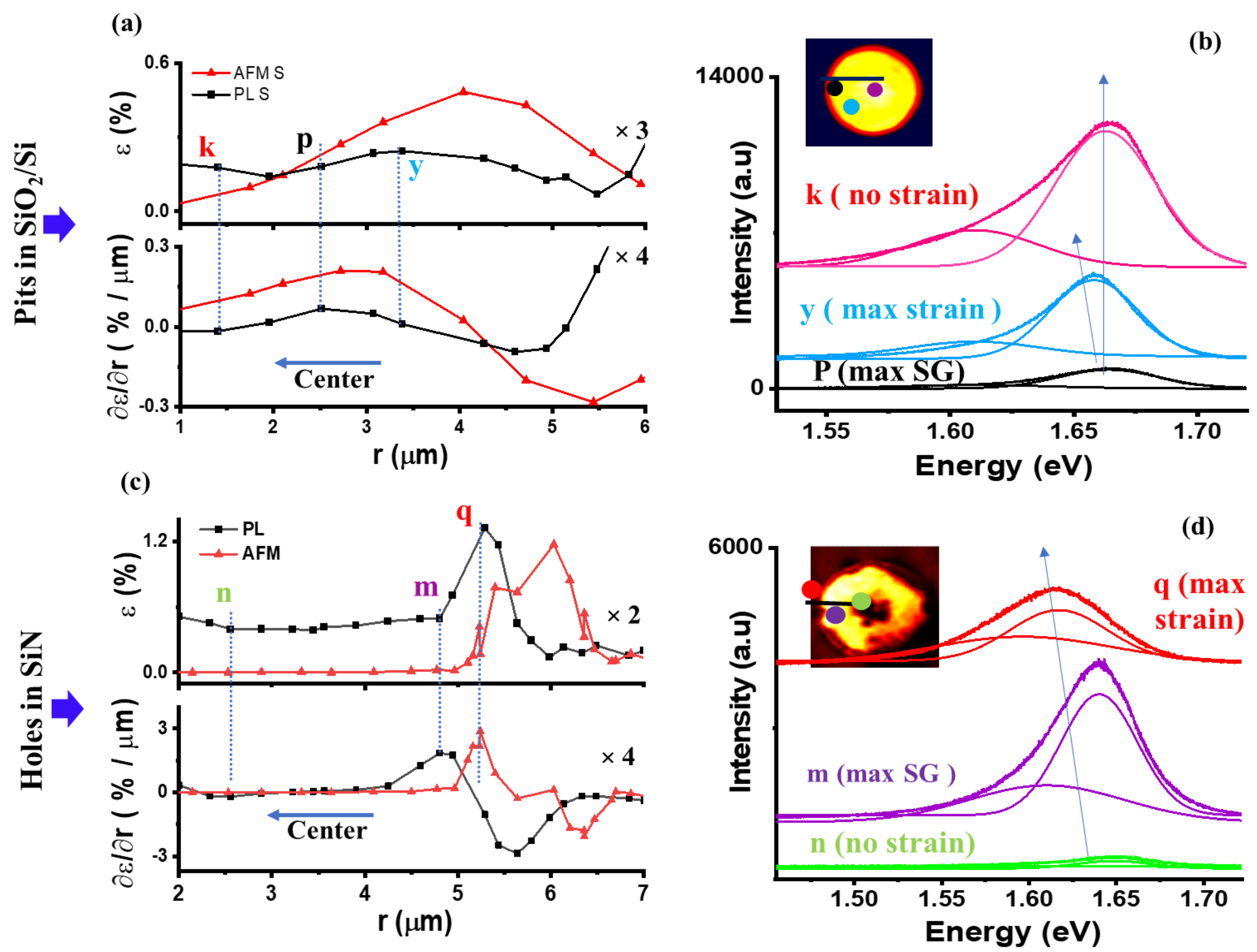

**Fig. S11 Comparison of strain and strain gradient from AFM and PL spectra.** (a) Strain and SG from AFM and PL map of the $10\mu m$ pit. (b) PL spectra at highest strain (y), highest SG (P) and at negligible strain (K) of the $10\mu m$ pit. (c) Strain and SG from AFM and PL map of the $10\mu m$ hole. (d) PL spectra at highest strain (q), highest SG (m) and at negligible strain (n) of the $10\mu m$ hole.

Fig. S11a shows the comparison of strain and SG computed from both AFM and PL data for the $10\mu m$ pit. For comparison the AFM data is down sampled to match the PL scan step size. The corresponding PL spectra is shown in Fig. S11b. As the SG is poor in the case of the pit, it fails to produce any funneling. Similar comparison of strain and SG of the hole is shown in Fig. S11c and the corresponding spectra in Fig. S11d. The point in the $m$ region shows maximum SG, therby maximum enhancement with moderate PL shift. The point in $q$ shows maximum PL shift with moderate enhancement as the strain is maximized but the SG is low. The PL data shows less strain w.r.t the AFM measurements. When we move from the center towards the circumference both the AFM and PL data shows that the SG peaks first and then the strain, this phase difference is due to fact that low strain area can give rise to high SG because the curvature is sharper. Also, we can see the SG alters sign both in the pit and in hole. This is because in case of the pit the direction of the SG alters near the circumference and in case of the hole the trench alters the SG direction. The lateral mismatch between the AFM and the PL plots is due to the different resolution between the measurements.